\PassOptionsToPackage{table,xcdraw}{xcolor}
\documentclass[11pt,a4paper]{article}
\pdfoutput=1
\usepackage{jheppub}
\usepackage{gensymb}
\DeclareSymbolFont{matha}{OML}{txmi}{m}{it}
\DeclareMathSymbol{\varv}{\mathord}{matha}{118}
\usepackage{amsmath}
\usepackage{amssymb}
\usepackage{graphicx}
\usepackage{subfigure}
\usepackage{pstricks}
\usepackage{bm}
\usepackage{pbox}
\usepackage{placeins}
\usepackage{booktabs}
\usepackage[T1]{fontenc}
\usepackage{footnote}
\usepackage{pdfpages}
\usepackage{hhline}
\usepackage{multirow}
\usepackage{multicol}
\usepackage{enumitem}
\usepackage[toc,page]{appendix}
\allowdisplaybreaks
\usepackage{comment}
\usepackage{float}
\usepackage{slashed}
\usepackage{xcolor}
\usepackage{textcomp}
\usepackage{gensymb}
\usepackage{multirow}
\usepackage[numbers]{natbib}
\usepackage{notoccite}

\usepackage{rotating}
\usepackage{tabularx}
\usepackage[labelfont=bf]{caption} 
   
\newcommand{\tc}{\textcolor}

\renewcommand{\thetable}{\Roman{table}}
\newcommand\scalemath[2]{\scalebox{#1}{\mbox{\ensuremath{\displaystyle #2}}}}

\definecolor{MyDarkBlue}{rgb}{0.1, 0.1, 0.8} 
\definecolor{MyLightBlue}{rgb}{0.22,0.51,0.9}
\definecolor{MyGreen}{rgb}{0.0, 0.5, 0.0}
\definecolor{BrickRed}{rgb}{0.8, 0.25, 0.33}

\RequirePackage{hyperref}
\hypersetup{colorlinks, citecolor=MyGreen,linkcolor=cyan, urlcolor=MyLightBlue}

\def\a{\alpha}

\def\d{\delta}

\def\k{\kappa}

\def\m{\mu}

\def\ve{\varepsilon}

\def\be{\begin{equation}}
\def\ee{\end{equation}}

\def\ra{\rightarrow}

\def\bea{\begin{eqnarray}}
\def\eea{\end{eqnarray}}

\definecolor{MyDarkBlue}{rgb}{0.1, 0.1, 0.8} 
\definecolor{MyLightBlue}{rgb}{0.22,0.51,0.9}
\definecolor{MyGreen}{rgb}{0.0, 0.5, 0.0}
\definecolor{BrickRed}{rgb}{0.8, 0.25, 0.33}

\RequirePackage{hyperref}
\hypersetup{colorlinks, citecolor=blue,linkcolor=cyan, urlcolor=MyLightBlue}

\makeatletter
\gdef\@fpheader{}
\makeatother
\begin{document}
\title{\bf 
Leptogenesis in SO(10) with Minimal Yukawa sector
}

\author[a]{K.S. Babu,}
\author[b]{Pasquale Di Bari,}
\author[c]{Chee Sheng Fong,}
\author[d]{Shaikh Saad}  

\vspace{0.5cm}

\affiliation[a]{Department of Physics, Oklahoma State University, Stillwater, OK 74078, USA}

\affiliation[b]{School of Physics and Astronomy, University of Southampton,
Southampton, SO17 1BJ, U.K.}

\affiliation[c]{Centro de Ciências Naturais e Humanas,
Universidade Federal do ABC, 09.210-170,\\ Santo André, SP, Brazil}

\affiliation[d]{Department of Physics, University of Basel, Klingelbergstrasse\ 82, CH-4056 Basel, Switzerland}

\emailAdd{babu@okstate.edu, P.Di-Bari@soton.ac.uk, sheng.fong@ufabc.edu.br, shaikh.saad@unibas.ch}
\abstract{
In prior studies,  a very minimal Yukawa sector within the $SO(10)$ Grand Unified Theory framework has been identified, comprising of  Higgs fields belonging to a real $10_H$, a real $120_H$, and a $\overline{126}_H$ dimensional representations. In this work, within this minimal framework, we have obtained fits to fermion masses and mixings while successfully reproducing the  cosmological baryon asymmetry via leptogenesis. 
The right-handed neutrino  ($N_i$) mass spectrum obtained from the fit is strongly hierarchical, suggesting that $B-L$ asymmetry is dominantly produced from $N_2$ dynamics while $N_1$ is responsible for erasing the excess asymmetry. With this rather constrained Yukawa sector, fits are obtained both for normal and inverted ordered neutrino mass spectra, consistent with leptonic CP-violating phase  $\delta_\mathrm{CP}$ indicated by global fits of neutrino oscillation data, while also satisfying the current limits from neutrinoless double beta decay experiments. In particular, the the leptonic CP-violating phase has a preference to be in the range $\delta_\mathrm{CP}\simeq (230-300)^\circ$. We also show the consistency of the framework with gauge coupling unification and proton lifetime limits.
}
\maketitle

\section{Introduction}
The Standard Model (SM) of particle physics is incredibly successful in describing the fundamental particles and their interactions. However, it does have some limitations and drawbacks. The model does not account for dark matter, it fails to incorporate 
the observed matter-antimatter asymmetry, and 
it predicts neutrinos to be massless, in contradiction with experiments. Several parameters of the model,  such as fermion masses, mixing angles and the number of families, are adjusted to experimental observations and lack a fundamental explanation, leading to the well-known flavor puzzle. 
 
Addressing these limitations is a significant focus of modern theoretical physics, with various models beyond the SM aiming at explaining  
some or all of these shortcomings.
Grand Unified Theories (GUTs)~\cite{Pati:1974yy, Georgi:1974sy, Georgi:1974yf, Georgi:1974my, Fritzsch:1974nn}   are well-motivated extensions which attempt such an explanation.
GUTs unify three of the fundamental forces of nature,  characterized by the SM gauge groups $SU(3)_c \times SU(2)_L \times U(1)_Y$ into a single force arising from a simple group such as $SU(5)$ and $SO(10)$. GUTs also unify quarks and leptons~\cite{Pati:1973rp}  into common multiplets and explain the small neutrino masses via the seesaw mechanism~\cite{Minkowski:1977sc,Yanagida:1979as,Glashow:1979nm,Gell-Mann:1979vob,Mohapatra:1979ia,Schechter:1980gr,Schechter:1981cv}.  
Furthermore, GUTs shine in 
simplicity and elegance by reducing the number of effective parameters through which  non-trivial correlations are obtained among the observed fermion masses and mixing angles. Finally, GUTs make testable predictions, such as proton decay,  which provide concrete ways to confront these theories.

Among possible GUT gauge groups, $SO(10)$ is arguably the most attractive candidate since it unifies all fermions of a generation into a single spinorial 16-dimensional representation. Additionally, the spinorial representation contains the right-handed neutrino, which, via the seesaw mechanism, provides the desired tiny masses to the light neutrinos. Remarkably, in the decay of these same right-handed neutrinos to leptons and the SM Higgs, an asymmetry in the lepton number is generated, a process termed leptogenesis~\cite{Fukugita:1986hr}, which through  electroweak sphaleron processes~\cite{Klinkhamer:1984di,Arnold:1987mh,Arnold:1987zg} gets partially converted into baryon asymmetry-- hence explaining the observed
matter-antimatter asymmetry of the Universe. 
(For reviews on leptogenesis, see for example Refs.  \cite{Buchmuller:2004nz,Nir:2007zq,Davidson:2008bu,Pilaftsis:2009pk,DiBari:2012fz,Fong:2012buy,Chun:2017spz,Dev:2017trv, Bodeker:2020ghk}.) Moreover, it turns out that the Yukawa sector of  $SO(10)$ GUTs  can be very predictive, which has been extensively analyzed in the literature~\cite{Babu:1992ia, Bajc:2001fe,Bajc:2002iw,Fukuyama:2002ch,Goh:2003sy,
Goh:2003hf,Bertolini:2004eq, Bertolini:2005qb, Babu:2005ia,Bertolini:2006pe, Bajc:2008dc,
Joshipura:2011nn,Altarelli:2013aqa,Dueck:2013gca, Fukuyama:2015kra, Babu:2016cri, Babu:2016bmy, Babu:2018tfi, Babu:2018qca, Ohlsson:2018qpt, Ohlsson:2019sja,Babu:2020tnf,Mummidi:2021anm,Saad:2022mzu,Haba:2023dvo, Kaladharan:2023zbr} to study correlations among quark and lepton masses and mixings as well as neutrino oscillation parameters. 

The focus of this paper is leptogenesis in
a class of $SO(10)$ GUTs with a very minimal Yukawa sector 
\emph{without assuming additional symmetries.} 
Symmetries exterior to $SO(10)$, such as flavor $U(1)$, Peccei-Quinn $U(1)$ and CP symmetry have been utilized in other studies in order to reduce the number of parameters.  The Yukawa couplings of the models that we study here is  based only on $SO(10)$ symmetry and involve  scalar fields belonging \emph{real} $10_H$, a real $120_H$ and a complex $126_H$~\cite{Babu:2016bmy}. Such a system, which contains 15 moduli and 12 phases, has been shown to result in realistic fermion masses and mixings.  Here we show that this setup can also generate the right amount of matter--antimatter asymmetry of the Universe via leptogenesis.

While light neutrino mass can arise from both type-I and type-II seesaw mechanisms, we focus here on the type-I dominated scenario (a pure type-II scenario, however, is incompatible with the observed fermion spectrum  in this setup~\cite{Babu:2016bmy}). Due to the very constraining Yukawa sector, a good fit to fermion observables predicts a hierarchical mass spectrum of right-handed neutrinos $N_i\,(i=1,2,3)$ with respective masses $\left(M_1, M_2, M_3\right)\sim     \left(10^{4-5}, 10^{11-12}, 10^{14-15}\right)\;\textrm{GeV}$. This is  in accordance with the expectations of various $SO(10)$-inspired models
\cite{Buchmuller:1996pa,Nezri:2000pb,Buccella:2001tq,Branco:2002kt,Akhmedov:2003dg,DiBari:2008mp,DiBari:2010ux,Buccella:2012kc,DiBari:2013qja,DiBari:2014eya,DiBari:2015oca,DiBari:2017uka,Chianese:2018rnq,DiBari:2020plh}.
This results in an interesting scenario where the final $B-L$ asymmetry is dominantly generated by the 
decays of the next-to-lightest right-handed neutrino, $N_2$, while the lightest right-handed neutrino, $N_1$, inverse decays 
wash-out the asymmetry at lower temperatures~\cite{DiBari:2005st}. 
Thanks to flavor effects, the $N_1$-wash-out is strongly reduced~\cite{Vives:2005ra} and in our case
a sizeable part of the asymmetry produced from $N_2$-decays can survive the $N_1$ wash-out in the muon flavor.
We show that this  in  accordance with the general analytical results from $SO(10)$-inspired leptogenesis scenarios \cite{DiBari:2008mp}.

Within the framework of  $SO(10)$ GUTs, baryogenesis via leptogenesis has been discussed in a series of works. In the $SO(10)$-inspired scenarios studied in Ref.~\cite{Buchmuller:1996pa,Nezri:2000pb,Buccella:2001tq,Branco:2002kt,Akhmedov:2003dg,DiBari:2008mp,DiBari:2010ux,Buccella:2012kc,DiBari:2013qja,DiBari:2014eya,DiBari:2015oca,DiBari:2017uka,Chianese:2018rnq,DiBari:2020plh} leptogenesis is analyzed in a general framework without having a predictive fermion spectrum.  The fermion mass spectrum and leptogenesis has been studied in tandem in Refs.~\cite{Fong:2014gea,Mummidi:2021anm,Patel:2022xxu,Kaladharan:2023zbr}. 
In particular, Refs.~\cite{Fong:2014gea,Mummidi:2021anm,Kaladharan:2023zbr} have considered leptogenesis in $SO(10) \times U(1)_{PQ}$ models with  the Yukawa sector consisting of scalar fields belonging to a complex $10_H$ and a $126_H$. This model gives rise to a more compact spectrum of right-handed neutrinos where their masses fall within the range $M_i \sim (10^9 - 10^{12})$ GeV and leptogenesis can have relevant contributions from decays of some or all $N_i$. On the other hand, Ref.~\cite{Patel:2022xxu} has considered leptogenesis with the same scalar field content but with $SO(10) \times {\rm CP}$  where the discrete CP symmetry makes all Yukawa couplings to be real, thus reducing the number of parameters. In such a  model, the spectrum of right-handed neutrinos is strongly hierarchical with $M_1 \sim 10^7$ GeV $\ll M_2 \ll M_3 \sim 10^{12}$ GeV such that leptogenesis proceeds mainly through decays of $N_2$ while $N_1$ erases part of the asymmetry generated. In contrast to these works, the present work assumes no additional symmetries beyond $SO(10)$ gauge symmetry.

By performing a detailed numerical analysis, 
we show that leptogenesis is viable for both normal and inverted mass ordering of light neutrinos. In particular, the value of the leptonic Dirac CP phase $\delta_{\rm CP}$ is consistent, both for normal and inverted mass ordering,  with the current global fits, which  mildly prefer nonzero  $\delta_{\rm CP}$~\cite{NUFIT}. 
Since we  aim for precision calculation, we utilize lepton-flavor-covariant formalism (i.e. independent of lepton flavor basis) developed in Ref.~\cite{Fong:2021xmi} which takes into account both lepton flavor effects~\cite{Dev:2017trv} and spectator effects~\cite{Buchmuller:2001sr,Nardi:2005hs}. The renormalization-group running effect is partially taken into account by considering two sets of Yukawa couplings, one at the scale $M_2$, the mass scale of $N_2$, and the other at the scale $M_1$
with the appropriate matching conditions imposed.  While the model admits both normal ordering (NO) and inverted ordering (IO) of neutrino masses, recent global fits and cosmological constraints show a preference to NO, which we analyze in more detail. In this case, the leptonic CP-violating phase is found, in our fit, to have a preferred value in the range $\delta_\mathrm{CP}\simeq (230-300)^\circ$. The best fit values of the parameters that explain simultaneously the fermion mass spectrum and the baryon asymmetry are listed in Table. \ref{result}.  

Our analysis reveals that for gauge coupling unification, it is crucial to incorporate some amount of threshold corrections to achieve large enough unification scale to be compatible with proton lifetime limits. Without these threshold corrections which arise from scalar sub-multiplets, proton lifetime would be too rapid and inconsistent with the current experimental bounds. A dedicated analysis is carried out to show the consistency of the model with gauge coupling unification and proton lifetime limits. Interestingly, the model prefers the lifetime to be on the lower end, which experiments such as Hyper-Kamiokande are expected to probe.

The rest of the article is organized as follows. In Sec.~\ref{sec:yukawa}, we introduce the model, and in Sec.~\ref{sec:lepto}, we provide all the necessary details to perform leptogenesis computation. In Sec.~\ref{sec:fit}, we undertake an extensive numerical analysis in search of finding a consistent fits of the fermion masses and mixings as well as reproducing the correct baryon asymmetry parameters. In Sec.~\ref{sec:SO10inspired} we provide some insights into the calculation
of the asymmetry, comparing our numerical results to the analytical results from $SO(10)$-inspired leptogenesis
\cite{DiBari:2014eya,DiBari:2017uka,DiBari:2020plh}. A detailed analyses of gauge coupling unification and proton decay is carried out in Sec.~\ref{sec:GCU}. 
Finally, we conclude in Sec.~\ref{sec:conclusions}.

\section{Minimal Yukawa sector}\label{sec:yukawa}
Assuming that there are no additional fermions  beyond the three families of chiral spinorial representation of $SO(10)$, the following fermion bilinear specifies the possible Higgs representations that can contribute to the fermion mass generation in a renormalizable theory: 
\begin{equation}
16\times 16 = 10_s+ 120_a + 126_s.
\end{equation}
Here the subscripts $s$ and $a$ represent symmetric and antisymmetric components in family  space. Note that $10$- and $120$-dimensional representations are real, whereas $126$-dimensional representation is inherently complex. With the above bilinears, the most general Yukawa sector of renormalizable $SO(10)$ GUTs can be written as 
\begin{equation}
\label{yukawa}
\mathcal{L}_{yuk}= 16_F (y_{10}^p10^p_H+y^q_{120}120^q_H+y_{126}^r\overline{126}^r_H) 16_F.
\end{equation}
In the above equation, we have suppressed the family index. If $n_{10}$ copies of $10$-dimensional representations are present, then the index $p$ takes values $p=1,2,...,n_{10}$; similarly  $q=1,2,...,n_{120}$ and $r=1,2,...,n_{126}$. Each Yukawa coupling $Y_{10}$ (and $Y_{126}$) is a symmetric $3\times 3$ matrix in the family space, whereas  $Y_{120}$ is antisymmetric.

It turns out that viable fermion spectrum can only be generated provided for $n_{10}+n_{120}+n_{126}>2$. 
By considering various possibilities, Ref.~\cite{Babu:2016bmy} identified that the most minimal Yukawa sector corresponds to $(n_{10},n_{120},n_{126})=(1,1,1)$. In this scenario, one has two symmetric and one antisymmetric Yukawa matrices, leading to a total of 3 real plus 9 complex Yukawa parameters.  Note that either $y_{10}$ or $y_{126}$ can be made real and diagonal using an $SO(10)$ rotation, which has been used for the parameter counting.  If any other realistic choice is considered, for example, $(n_{10},n_{120},n_{126})=(2,0,1)$, then one would end up with larger number of parameters, viz., 3 real and 12 complex Yukawa parameters for this choice.

There is an alternative option to reduce the number of Yukawa parameters, which however, requires extending the symmetry of the theory. Along this line, the most studied case in the literature is $(n_{10},n_{120},n_{126})=(2,0,1)$, where two 10-dimensional representations are complexified using a Peccei-Quinn $U(1)_\mathrm{PQ}$. In this construction, only one of the two Yukawa couplings associated to $10_H$ is allowed due to the PQ symmetry, thereby reducing 6 complex Yukawa parameters~\cite{Babu:1992ia}. 

In this work, we stick to the scenario where the full symmetry of the theory is nothing but $SO(10)$ gauge symmetry, and consider the most minimal Yukawa sector that corresponds to $(n_{10},n_{120},n_{126})=(1,1,1)$. Then, the up-type quark, down-type quark, charged leptons, Dirac neutrino, and Majorana neutrino mass matrices can be written as~\cite{Babu:2016bmy}:
\begin{align}
&M_U= \underbrace{v_{10}y_{10}}_{\equiv D}+\underbrace{v^u_{126}y_{126}}_{\equiv S}+\underbrace{(v^{(1)}_{120}+v^{(15)}_{120})y_{120}}_{\equiv A},\label{Mu}\\
&M_D= v^{\ast}_{10}y_{10}+v^d_{126}y_{126}+(v^{(1)\ast}_{120}+v^{(15)\ast}_{120})y_{120},\\
&M_E= v^{\ast}_{10}y_{10}-3v^d_{126}y_{126}+(v^{(1)\ast}_{120}-3v^{(15)\ast}_{120})y_{120},\\
&M_{\nu_D}= v_{10}y_{10}-3v^u_{126}y_{126}+(v^{(1)}_{120}-3v^{(15)}_{120})y_{120},\\
&M_{\nu_{R}}=v_{R}y_{126}. \label{MR}
\end{align}
There are a few special features of the above set of mass matrices. For example, the same VEV (vacuum expectation value) $v_{10}$ enters in both the up-sector (up-type quark and Dirac neutrino) as well as in the down-sector (down-type quark and charged lepton). This happens because $10_H$ is a real representation and it contains a self-conjugate bi-doublet, $(1,2,2)$, under the $SU(4)_c \times SU(2)_L \times SU(2)_R$ (Pati-Salam) subgroup of $SO(10)$.   Consequently, the reality of $10_H$ implies that $v_u=v_d^*\equiv v_{10}$. Additionally, the reality of $120_H$ implies, $v^{(1)}_u=v^{(1)*}_d\equiv v^{(1)}_{120}$ and $v^{(15)}_u=v^{(15)*}_d \equiv v^{(15)}_{120}$. Here $v^{(1)}_{120}$ and $v^{(15)}_{120}$ 
represent the VEVs of the submultiplets $(1,2,2)$ and $(15,2,2)$ contained in $120_H$, under the Pati-Salam symmetry. The VEVs of the up-type and down-type weak doublets contained in $(15,2,2)$ arising from the complex  $126_H$, which we denote by $v^u_{126}$ and $v^d_{126}$, have no such relations. 
We now define
\begin{align}
r_1&=\frac{v^d_{126}}{v^u_{126}},\;\; r_2=\frac{v^{(1)\ast}_{120}-3v^{(15)\ast}_{120}}{v^{(1)}_{120}+v^{(15)}_{120}},\;\; e^{i\phi}= \frac{v^{(1)\ast}_{120}+v^{(15)\ast}_{120}}{v^{(1)}_{120}+v^{(15)}_{120}},\;\; c_{R}=\frac{v_{R}}{v^u_{126}},\label{cR}
\end{align}
and rewrite the above mass matrices as
\begin{align}
&M_{U}= \underbrace{D}_\mathrm{symetric}+\underbrace{S}_\mathrm{real-diagonal} +\underbrace{A}_\mathrm{antisymetric} \equiv vy_U, \label{E1}\\
&M_{D}= D+r_1 S+ e^{i \phi} A\equiv vy_D, \\
&M_{E}= D-3 r_1 S+ r_2 A\equiv vy_E, \label{ME} \\
&M_{\nu_D}= D-3 S+ r^{\ast}_2 e^{i \phi} A\equiv vy_{\nu_D}, \label{MDnu}\\
&M_{\nu_{R}} = c_{R} S , \label{E2}
\end{align}
with $v^2 = |v_{10}|^2 + |v_{120}^{(1)}|^2 + |v_{120}^{(15)}|^2 + |v_{126}^u|^2 + |v_{126}^d|^2$, and $v=174.104$ GeV. 
These Yukawa coupling matrices, $y_f$, (with $f=U, D, E, \nu_D$) are matched to the following Lagrangian of the SM extended with right-handed neutrinos:
\begin{align}
-\mathcal{L}_{yuk}&=(y_{\nu_D})_{ij} \overline {N_i}\ell_j\epsilon H+(y_U)_{ij}\overline {U_i}Q_j\epsilon H
+(y_D)_{ij}\overline {D_i}Q_jH^*+(y_E)_{ij}\overline {E_i} \ell_j H^* 
\nonumber\\&
+ \frac{1}{2}(M_{\nu_R})_{ii}\overline {N_i} N^c_i + \textrm{h.c.},
\label{eq:yukawa_terms}
\end{align}
where $Q_j$, $\ell_j$ and $H$ are the left-handed quark, lepton and Higgs $SU(2)_L$ doublets, respectively; $U_i$, $D_i$, $E_i$ and $N_i$ are the right-handed up-type quark, down-type quark, charged lepton and neutrino $SU(2)_L$ singlets, respectively, $i,j=1,2,3$ are the family indices, the superscript $c$ denotes charge conjugation and $\epsilon$ is the antisymmetric tensor for $SU(2)_L$ contraction.

Here we have chosen a phase convention where $v_{10}$ is made real by an $SU(2)_L$ rotation. 
Without loss of generality, we choose to work in a basis where the matrix $S$ is diagonal and real, as specified above. When the charged fermion Yukawa coupling matrices are diagonalized, the corresponding diagonal couplings are denoted by  $y^\textrm{diag}_U=\left(y_u,y_c,y_t\right)$, $y^\textrm{diag}_D=\left(y_d,y_s,y_b\right)$, and $y^\textrm{diag}_E=\left(y_e,y_\mu,y_\tau\right)$. 
The light neutrino mass matrix $m_\nu$ for the SM neutrinos $\nu_i$, obtained from the seesaw formula~\cite{Minkowski:1977sc,Yanagida:1979as,Glashow:1979nm,GellMann:1980vs,Mohapatra:1979ia}, in the basis $\overline {\nu^c}\, m_\nu\, \nu/2$  
is given by,
\begin{align}
\label{eq:mnu}
m_\nu = - M^T_{\nu_D} M^{-1}_{\nu_R} M_{\nu_D} ,
\end{align}
where we have assumed the type-I dominance.  We diagonalize the neutrino mass matrix as follows:
\begin{align}
    m_\nu = N_L^* \mathrm{diag}(m_1,m_2,m_3)N_L^\dagger ,
\end{align}
where $N_L$ is a unitary matrix.
Then, the PMNS matrix is defined as
\begin{align}
U_\nu = E_L^\dagger N_L,    
\end{align}
where the unitary matrix $E_L$ is obtained by diagonalizing the charged lepton mass matrix $M_E=E_R \,\mathrm{diag}(m_e,m_\mu,m_\tau) E_L^\dagger$ where $E_R$ is the right-handed analog of $E_L$. We parametrize the Pontecorvo–Maki–Nakagawa–Sakata (PMNS) matrix as:
\begin{align}
U_\nu = &   
\left( \begin{array}{ccc}
c_{12}\,c_{13} & s_{12}\,c_{13} & s_{13}\,e^{-{\rm i}\,\d_\mathrm{CP}} \\
-s_{12}\,c_{23}-c_{12}\,s_{23}\,s_{13}\,e^{{\rm i}\,\d_\mathrm{CP}} &
c_{12}\,c_{23}-s_{12}\,s_{23}\,s_{13}\,e^{{\rm i}\,\d_\mathrm{CP}} & s_{23}\,c_{13} \\
s_{12}\,s_{23}-c_{12}\,c_{23}\,s_{13}\,e^{{\rm i}\,\d_\mathrm{CP}}
& -c_{12}\,s_{23}-s_{12}\,c_{23}\,s_{13}\,e^{{\rm i}\,\d_\mathrm{CP}}  &
c_{23}\,c_{13}
\end{array}\right)
\nonumber \\
&
\times \mathrm{diag}(e^{-i\alpha/2},e^{-i\beta/2},1), 
\label{Upmns-convention}
\end{align}
where $c_{ij} = \cos \theta_{ij}$ and $s_{ij} = \sin\theta_{ij}$.

The number of parameters contained in the fermion mass matrices given in Eqs.~\eqref{E1}-\eqref{E2} may be counted as follows. Matrices $S$, $A$, and $D$ contain 3 real, 3 complex, and 6 complex parameters, respectively.  The ratios $r_{1,2}$ are complex, whereas, the phase of $c_R$ is irrelevant for fermion mass fit. Moreover, there is a non-trivial phase, $\phi$, as defined in Eq.~\eqref{cR}.  Altogether, we have 15 real parameters and 12 phases. If all parameters were real, then these 15 real parameters would determine 17 observables, namely, 9 charged fermion masses, 3 quark mixing angles, 2 neutrino mass-squared differences, and 3 neutrino mixing angles.  Such a system, however, would not explain the observed  CP violation in the quark sector, and also would not lead to a realistic fermion spectrum. Hence the phases play a crucial role in providing a correct fit to the masses and mixings. Since 12 of the parameters are phases, it is nontrivial to achieve a consistent fit.

\section{Leptogenesis}\label{sec:lepto}
Aiming to determine the production of baryon asymmetry from leptogenesis as precise as possible, we will utilize the lepton-flavor-covariant formalism developed in Ref.~\cite{Fong:2021xmi} which includes both lepton flavor effects~\cite{Dev:2017trv} and spectator effects~\cite{Buchmuller:2001sr,Nardi:2005hs}. 
The renormalization-group-running effects will also be partially taken into account as described at the end of this section.
First we review the formalism used in this work.

Let us define the number asymmetry of particle species $i$ as
\begin{eqnarray}
Y_{\Delta i} & \equiv Y_{i}-Y_{\bar{i}}= & \frac{n_{i}}{s}-\frac{n_{\bar{i}}}{s},
\end{eqnarray}
where $n_{i(\bar{i})}$ is particle (antiparticle) number density
of species $i$ and $s=\frac{2\pi^{2}}{45}g_{\star}T^{3}$ is the cosmic entropy
density and $g_{\star}$ is the effective relativistic degrees of
freedom of the Universe at the relevant temperature. The baryon number charge ($B$) can be constructed as
\begin{eqnarray}
Y_{B} & = & \sum_{i}q_{i}^{B}Y_{\Delta i},
\end{eqnarray}
where $q_{i}^{B}$ is the baryon number carried by species $i$. 

Since we are interested in taking into account lepton flavor effect\tc{brown}{s},
we will consider $3\times3$ matrices of number asymmetries of lepton
doublets $\ell$ and charged lepton singlet $E$, $Y_{\Delta\ell}$
and $Y_{\Delta E}$
\begin{eqnarray}
Y_{\Delta\ell} & = & \left(\begin{array}{ccc}
Y_{\Delta\ell_{11}} & Y_{\Delta\ell_{12}} & Y_{\Delta\ell_{13}}\\
Y_{\Delta\ell_{12}}^{*} & Y_{\Delta\ell_{22}} & Y_{\Delta\ell_{23}}\\
Y_{\Delta\ell_{13}}^{*} & Y_{\Delta\ell_{23}}^{*} & Y_{\Delta\ell_{33}}
\end{array}\right),\;\;\;\;\;Y_{\Delta E}=\left(\begin{array}{ccc}
Y_{\Delta E_{11}} & Y_{\Delta E_{12}} & Y_{\Delta E_{13}}\\
Y_{\Delta E_{12}}^{*} & Y_{\Delta E_{22}} & Y_{\Delta E_{23}}\\
Y_{\Delta E_{13}}^{*} & Y_{\Delta E_{23}}^{*} & Y_{\Delta E_{33}}
\end{array}\right),
\end{eqnarray}
where the basis-independent quantities $\textrm{Tr}\left(Y_{\Delta\ell}\right)$
and $\textrm{Tr}\left(Y_{\Delta E}\right)$ are respectively the total
number asymmetries of lepton doublets $\ell$ and charged lepton singlet $E$. The off-diagonal elements are needed to describe lepton flavor coherence and correlations.

Let us further define the following baryon charge matrix
\begin{eqnarray}
Y_{\widetilde{\Delta}} & \equiv & \frac{1}{3}Y_{B}I_{3\times3}-Y_{\Delta\ell}.\label{eq:tildeDelta_matrix}
\end{eqnarray}
The evolution of $Y_{\widetilde{\Delta}}$ due to the SM charged lepton Yukawa interactions can be described by
\begin{eqnarray}
s{\cal H}z\frac{dY_{\widetilde{\Delta}}}{dz} & = & \frac{\gamma_{E}}{2Y^{{\rm nor}}}\left\{ y_{E}^{\dagger}y_{E},\frac{Y_{\Delta\ell}}{g_{\ell}\zeta_{\ell}}\right\} -\frac{\gamma_{E}}{Y^{{\rm nor}}}y_{E}^{\dagger}y_{E}\frac{Y_{\Delta H}}{g_{H}\zeta_{H}}-\frac{\gamma_{E}}{Y^{{\rm nor}}}y_{E}^{\dagger}\frac{Y_{\Delta E}}{g_{E}\zeta_{E}}y_{E} \label{eq:BE_YtildeDelta}
\end{eqnarray}
where we have defined the anticommutator $\left\{ \mathcal{A},\mathcal{B}\right\} \equiv \mathcal{AB+BA}$,
$z\equiv M_{\textrm{ref}}/T$ with $M_{\textrm{ref}}$ an arbitrary
mass scale, ${\cal H}=1.66\sqrt{g_{\star}}T^{2}/M_{\textrm{Pl}}$ with $M_{\rm Pl} = 1.22\times 10^{19}$ GeV is the Hubble rate during the radiation dominated era, $Y^{{\rm nor}}\equiv\frac{15}{8\pi^{2}g_{\star}}$
, $g_{i}$ is the gauge degrees of freedom of species  $i$ (with $g_{\ell}=g_{H}=2$
and $g_{E}=1$) and
\begin{eqnarray}
\zeta_{i} & \equiv & \frac{6}{\pi^{2}}\int_{m_{i}/T}^{\infty}dx \, x\sqrt{x^{2}-m_{i}^{2}/T^{2}}\frac{e^{x}}{\left(e^{x}+\xi_{i}\right)^{2}},
\end{eqnarray}
with $m_{i}$ the mass of $i$ and $\xi_{i}=1(-1)$ for fermion (boson).
Since we are considering the condition before electroweak symmetry
breaking, we will take $m_{\ell}=m_{E}=m_{H}=0$ which results in
$\zeta_{\ell}=\zeta_{E}=\zeta_{H}/2=1$. The charged lepton Yukawa
reaction density $\gamma_{E}$ was determined in Refs.~\cite{Garbrecht:2013bia,Garbrecht:2014kda} to be $\gamma_{E}\approx5\times10^{-3}\,\frac{T^{4}}{6}$, which we shall make use of. 

Assuming that the electroweak sphaleron freezes out at 132 GeV after the electroweak phase transition temperature of 160 GeV~\cite{DOnofrio:2014rug}, we obtain the final baryon charge asymmetry as
\begin{eqnarray}
    Y_{B} = \left. c_{sp}(T)\textrm{Tr}(Y_{\tilde \Delta} - Y_{\Delta E})\right|_{T = 132\,\textrm{GeV}} \label{eq:B-L_to_B}
\end{eqnarray}
where assuming the SM degrees of freedom, we have~\cite{Fong:2015vna}
\begin{eqnarray}
    c_{sp}(T) = \frac{6(5 + \zeta_t)}{97 + 14 \zeta_t}.
\end{eqnarray}
The extreme values of this parameter are $c_{sp} = 12/37$ for $\zeta_t = 1$ and  $c_{sp} = 30/97$ for $\zeta_t = 0$.
Using the top mass $m_t = 173$ GeV, we have $c_{sp} = 0.315$ which is the value we will use. To convert to $\eta_B \equiv n_{B}/n_\gamma$ i.e. the baryon number density normalized to the photon number density today, we multiply $Y_{B}$ by the ratio of entropy to photon number density today $s_0/n_{\gamma 0} = $ 7.039.

From Eq.~\eqref{eq:yukawa_terms}, we can write down the relevant type-I seesaw Lagrangian with $M_i \equiv (M_{\nu_R})_{ii}$ and $y\equiv y_{\nu D}$ in the mass basis of $N_i$ as follows:
\begin{eqnarray}
-{\cal L} & \supset & \frac{1}{2}M_{i} \overline {N_{i}} N_{i}^{c}+y_{i\alpha} \overline{N_{i}}\ell_{\alpha}\epsilon H+\left(y_{E}\right)_{\alpha\beta}\overline{E_{\alpha}}\ell_{\beta}H^{*}+{\rm h.c.},\label{eq:type_I}
\end{eqnarray}
where lepton number $L$ or $B-L$ are explicitly broken, and for readability we have used Greek indices $\alpha,\beta = 1,2,3$ to denote the SM lepton flavors to distinguish them from the flavors of $N_i \, (i=1,2,3)$. 
The evolution of $Y_{N_{i}}$ is described by
\begin{equation}
s{\cal H}z\frac{dY_{N_{i}}}{dz}=-\gamma_{N_{i}}\left(\frac{Y_{N_{i}}}{Y_{N_{i}}^{{\rm eq}}}-1\right),
\end{equation}
where we now fix the reference scale by defining $z\equiv M_{1}/T$.
In our calculations, we assume that the $SO(10)$ GUT  is broken before/during the inflation and after inflation, the reheat temperature $T_R$ is below the GUT scale, so that the GUT gauge bosons are not brought into thermal equilibrium. We can estimate if $N_i$ will be thermalized through GUT gauge interaction by estimating the relevant rate by
\begin{eqnarray}
    \Gamma_{R} \sim \frac{T^5}{\pi^3 v_R^4}.
\end{eqnarray}
Comparing to the Hubble rate ${\cal H}(T)$ assuming radiation domination, $N_i$ will be thermalized if the cosmic temperature satisfies
\begin{eqnarray} \label{Ntherm}
    T > T_{eq} \sim 1.6 \times 10^{13}\,\textrm{GeV}
    \left(\frac{v_R}{10^{14}\,\textrm{GeV}}\right)^4,
\end{eqnarray}
where we have set $g_\star = 106.75$.
If $T_R < T_{eq}$ or $M_i > T_{eq}$, $N_i$ will not be thermalized. Since $T_R$ and also $v_R$ are unknown, we will quote the results for the baryon asymmetry assuming zero and thermal initial abundances of $N_i$ in the next section.

Next, to Eq. (\ref{eq:BE_YtildeDelta}), we introduce a source and
a washout term respectively given by\footnote{We will consider only decay and inverse decay of $N_{i}$ since scatterings
are in general suppressed.}
\begin{eqnarray}
S^{{\rm I}} & \equiv & -\sum_{i}\epsilon_{i}\gamma_{N_{i}}\left(\frac{Y_{N_{i}}}{Y_{N_{i}}^{{\rm eq}}}-1\right),\label{eq:source}\\
W^{{\rm I}} & \equiv & \frac{1}{2}\sum_{i}\frac{\gamma_{N_{i}}}{Y^{{\rm nor}}}\left(\frac{1}{2}\left\{ P_{i},\frac{Y_{\Delta\ell}}{g_{\ell}\zeta_{\ell}}\right\} +P_{i}\frac{Y_{\Delta H}}{g_{H}\zeta_{H}}\right).\label{eq:washout}
\end{eqnarray}
We will further assume Maxwell-Boltzmann distribution to calculate
the decay reaction density,
\begin{eqnarray}
\gamma_{N_{i}} & = & sY_{N_{i}}^{{\rm eq}}\Gamma_{N_{i}}\frac{{\cal K}_{1}\left(M_{i}/T\right)}{{\cal K}_{2}\left(M_{i}/T\right)},
\end{eqnarray}
where $Y_{N_{i}}^{{\rm eq}}=\frac{45}{2\pi^{4}g_{\star}}\frac{M_{i}^{2}}{T^{2}}{\cal K}_{2}\left(\frac{M_{i}}{T}\right)$
with ${\cal K}_{n}\left(x\right)$ the modified Bessel function of
the second kind of order $n$ and $\Gamma_{N_{i}}=\frac{\left(yy^{\dagger}\right)_{ii}M_{i}}{8\pi}$
is the total decay width of $N_{i}$. The CP parameter matrix $\epsilon_{i}$
and flavor projection matrix $P_{i}$ are respectively given by~\cite{Blanchet:2011xq}
\begin{eqnarray}
\left(\epsilon_{i}\right)_{\alpha\beta} & = & \frac{1}{16\pi}\frac{i}{\left(yy^{\dagger}\right)_{ii}}\sum_{j\neq i}\left[\left(yy^{\dagger}\right)_{ji}y_{j\beta}y_{i\alpha}^{*}-\left(yy^{\dagger}\right)_{ij}y_{i\beta}y_{j\alpha}^{*}\right]g\left(\frac{M_{j}^{2}}{M_{i}^{2}}\right)\nonumber \\
 &  & +\frac{1}{16\pi}\frac{i}{\left(yy^{\dagger}\right)_{ii}}\sum_{j\neq i}\left[\left(yy^{\dagger}\right)_{ij}y_{j\beta}y_{i\alpha}^{*}-\left(yy^{\dagger}\right)_{ji}y_{i\beta}y_{j\alpha}^{*}\right]\frac{M_{i}^{2}}{M_{i}^{2}-M_{j}^{2}},\label{eq:CP_typeI}\\
P_{i} & = & \frac{1}{\left(yy^{\dagger}\right)_{ii}}\left(\begin{array}{ccc}
\left|y_{i1}\right|^{2} & y_{i1}^{*}y_{i2} & y_{i1}^{*}y_{i3}\\
y_{i1}y_{i2}^{*} & \left|y_{i2}\right|^{2} & y_{i2}^{*}y_{i3}\\
y_{i1}y_{i3}^{*} & y_{i2}y_{i3}^{*} & \left|y_{i3}\right|^{2}
\end{array}\right),
\end{eqnarray}
where
\begin{equation}
    g(x) \equiv \sqrt{x}\left[\frac{1}{1-x}+1-(1+x)\ln\frac{1+x}{x}\right].
\end{equation}
 
Under arbitrary flavor rotations
\begin{eqnarray}
E & \to & U_E E,\quad\ell\to V_\ell \ell,\quad y \to y V_\ell^\dagger,\quad y_{E}\to U_E y_{E}V_\ell^{\dagger},\label{eq:flavor_rotations}
\end{eqnarray}
where $U_E$ and $V_\ell$ are unitary matrices, the density matrix of Eq. (\ref{eq:BE_YtildeDelta}), including the source
and washout terms of Eqs. (\ref{eq:source})-(\ref{eq:washout}), is manifestly
covariant with
\begin{eqnarray}
Y_{\Delta\ell} & \to & V_\ell Y_{\Delta\ell}V_\ell^{\dagger},\;\;\;\;\;Y_{\Delta E}\to U_E Y_{\Delta E}U_E^{\dagger},\label{eq:transformation_leptonic_charges}\\
\epsilon_{i} & \to & V_\ell \epsilon_{i}V_\ell^{\dagger},\;\;\;\;\;P_{i}\to V_\ell P_{i}V_\ell^{\dagger}.
\end{eqnarray}
From Eq.~\eqref{eq:B-L_to_B}, we see that the $B$ or $B-L$ asymmetry at any moment is invariant under such rotations.

To have closed set of equations, we can write~\cite{Fong:2021xmi}
\begin{eqnarray}
Y_{\Delta\ell} & = & \frac{2}{15}c_{B}{\rm Tr}Y_{\widetilde{\Delta}}-Y_{\widetilde{\Delta}},\label{eq:Yell_Ycharges}\\
Y_{\Delta H} & = & -c_{H}\left({\rm Tr}Y_{\widetilde{\Delta}}-2{\rm Tr}Y_{\Delta E}\right),\label{eq:YH_Ycharges}
\end{eqnarray}
where $Y_{\Delta\ell}$ is now a $3\times3$ matrix with off-diagonal
element $\alpha\neq\beta$ given by $\left(Y_{\Delta\ell}\right)_{\alpha\beta}=-\left(Y_{\widetilde{\Delta}}\right)_{\alpha\beta}$.
The temperature-dependent coefficients $c_{B}$ and $c_{H}$ can capture \emph{all} the spectator effects in the SM. Across $T_{B}\sim2\times10^{12}$ GeV,
the EW sphaleron interaction gets into equilibrium. To within percent
level precision, the following fitting function can be used~\cite{Fong:2020fwk}
\begin{eqnarray}
c_{B}\left(T\right) & = & 1-e^{-\frac{T_{B}}{T}},
\end{eqnarray}
where $T_{B}=2.3\times10^{12}$ GeV. The rest of the quark spectator
effects are described by $c_{H}$ with~\cite{Fong:2020fwk}
\begin{eqnarray}
c_{H}\left(T\right) & = & \begin{cases}
1 & T>T_{t}\\
\frac{2}{3} & T_{u}<T<T_{t}\\
\frac{14}{23} & T_{u-b}<T<T_{u}\\
\frac{2}{5} & T_{u-c}<T<T_{u-b}\\
\frac{4}{13} & T_{B_{3}-B_{2}}<T<T_{u-c}\\
\frac{3}{10} & T_{u-s}<T<T_{B_{3}-B_{2}}\\
\frac{1}{4} & T_{u-d}<T<T_{u-s}\\
\frac{2}{11} & T<T_{u-d}
\end{cases},\label{eq:cH}
\end{eqnarray}
where we use the transition temperatures $T_x$ given in Ref.~\cite{Fong:2021xmi}.
One can parametrize these transitions with the following function:
\begin{eqnarray}
c_{H}\left(T\right) & = & \left(\frac{2}{3}+\frac{1}{3}e^{-\frac{T_{t}}{T}}\right)-\left(\frac{2}{3}-\frac{14}{23}\right)\left(1-e^{-\frac{T_{u}}{T}}\right)-\left(\frac{14}{23}-\frac{2}{5}\right)\left(1-e^{-\frac{T_{u-b}}{T}}\right)\nonumber \\
 &  & -\left(\frac{2}{5}-\frac{4}{13}\right)\left(1-e^{-\frac{T_{u-c}}{T}}\right)-\left(\frac{4}{13}-\frac{3}{10}\right)\left(1-e^{-\frac{T_{B_{3}-B_{2}}}{T}}\right)\nonumber \\
 &  & -\left(\frac{3}{10}-\frac{1}{4}\right)\left(1-e^{-\frac{T_{u-s}}{T}}\right)-\left(\frac{1}{4}-\frac{2}{11}\right)\left(1-e^{-\frac{T_{u-d}}{T}}\right).
\end{eqnarray}

In the model under discussion, as we will see in the next section, due to the mass spectrum of $N_i$ imposed by the model, $N_2$ is mainly responsible for asymmetry generation while $N_1$ dynamics is mainly responsible for washout and the role of $N_3$ is important in yielding the necessary interference to have non-vanishing $C\!P$ violation in 
$N_2$ decays: it should be then appreciated how the existence of 
three families is crucial to have successful leptogenesis. 

For the calculation of the final asymmetry, we partially take into account  renormalization-group running by considering two sets of Yukawa couplings, one fixed at renormalization scale $\mu = M_2$ throughout $N_2$ leptogenesis while another is fixed at scale $\mu = M_1$ throughout $N_1$ washout. This running effect is important since the washout depends exponentially on the decay parameter defined as
\begin{eqnarray}
    K_i(\mu) \equiv \frac{\Gamma_{N_i}(\mu)}{{\cal H}(T=M_i)}
    \equiv \frac{(yy^\dagger)_{ii} v^2}{m_\star M_i},
    \label{eq:washout_parameter}
\end{eqnarray}
where $m_\star \equiv 1.66\sqrt{g_\star}\times 8\pi v^2/M_{\rm Pl}$ and $K_i >(<) 1$ is known as strong (weak) washout regime in which $N_i$ does (not) thermalize before decay.
The two sets of Yukawa couplings $y$ and $y_E$ in the right-handed neutrino mass basis at renormalization scale $\mu=M_2$ and $\mu=M_1$ are given in Appendix \ref{C}. 
One should have in mind that, despite a full account of
flavor and spectator effects, 
in the intermediate to strong washout regimes (as indicated by our benchmark points in the next section), there remains a theoretical uncertainty of ${\cal O}(1)$
since we are not taking into account full running of parameters, as well as thermal and next-to-leading order corrections~\cite{Salvio:2011sf,Laine:2011pq,Garbrecht:2019zaa}.

\section{Numerical analysis}\label{sec:fit}
In this section, we perform a combined fit to the fermion masses and mixings along with the baryon asymmetry of the universe. 
In our fitting procedure, we first use the approximate formulas for the calculation of baryon asymmetry through leptogenesis as in Ref.~\cite{Blanchet:2011xq} to identify candidate points which might produce sufficient baryon asymmetry. 
Only after a good fit is obtained, we carry out a full numerical calculation for baryon asymmetry using the formalism explained in the previous section to see if the resulting baryon asymmetry is consistent with the observed value.
The details of our fitting procedure are summarized  below.

\subsection{Benchmark fits}
The free parameters $S_{ii}, D^{j\geq i}_{ij}, A^{j> i}_{ij}$, $r_{1,2,}, c_R$, and $\phi$ are varied randomly at the GUT scale, which we choose to be $M_{\rm GUT}=2\times 10^{16}$ GeV. We then run the relevant Yukawa couplings defined in Eqs.~\eqref{E1}-\eqref{MDnu} and the right-handed neutrino mass matrix of Eq.~\eqref{E2} by taking  into account the full SM+Type-I seesaw renormalization group equations (RGEs) from $M_{\rm GUT}$ to $M_Z$ scale. In this running procedure, heavy right-handed neutrinos are successively integrated out at their respective mass thresholds. We have implemented our model in the package \texttt{REAP}~\cite{Antusch:2005gp} to take into account this running. Since the model under investigation predicts the intermediate symmetry breaking scale, $M_\mathrm{int}\sim 10^{14}$ GeV (as shown below), which is very close to the GUT scale, for our purpose, it is good enough to consider only  SM+Type-I RGEs below the GUT scale. Note that to be consistent with the present experimental bounds~\cite{Super-Kamiokande:2020wjk} on gauge-mediated proton decay, the GUT scale approximately needs to satisfy  $M_\mathrm{GUT} \gtrsim 5\times 10^{15}$ GeV. Consistency of obtaining such a large intermediate scale as well as evading proton decay bounds are discussed  in Sec.~\ref{sec:GCU}.

\FloatBarrier
\begin{table}[t!]
\centering
\small
\resizebox{0.9\textwidth}{!}{
\begin{tabular}{|c||c|c|c|}
\hline
\textbf{Observables} & \multicolumn{3}{c|}{Values at $M_Z$ scale}  \\ 

\cline{2-4}
($\Delta m^2_{ij}$ in ${
\rm eV}^2$) &Input&Benchmark Fit: NO& Benchmark Fit: IO 
\\
\hline \hline

\rowcolor{teal!20}$y_u/10^{-6}$&6.65$\pm$2.25&7.30&10.0\\  
\rowcolor{teal!20}$y_c/10^{-3}$&3.60$\pm$0.11&3.59&3.57\\ 
\rowcolor{teal!20}$y_t$&0.986$\pm$0.0086&0.986& 0.986\\ 
\hline\hline

\rowcolor{red!13}$y_d/10^{-5}$&1.645$\pm$0.165&1.636&1.635\\ 
\rowcolor{red!13}$y_s/10^{-4}$&3.125$\pm$0.165&3.122&3.148\\ 
\rowcolor{red!13}$y_b/10^{-2}$&1.639$\pm$0.015&1.639&1.637\\ 
\hline\hline

\rowcolor{blue!10}$y_e/10^{-6}$&2.7947$\pm$0.02794&2.7945&2.7906\\ 
\rowcolor{blue!10}$y_\mu/10^{-4}$&5.8998$\pm$0.05899&5.9011&5.9080\\ 
\rowcolor{blue!10}$y_\tau/10^{-2}$&1.0029$\pm$0.01002&1.0022&1.0023\\ 
\hline\hline

\rowcolor{yellow!13}$\theta_{12}^\textrm{CKM}/10^{-2}$&$22.735\pm$0.072&22.729 ($\theta_{12}^\textrm{CKM}=$13.023$^\circ$)&22.730 ($\theta_{12}^\textrm{CKM}=$13.023$^\circ$)\\ 
\rowcolor{yellow!13}$\theta_{23}^\textrm{CKM}/10^{-2}$&4.208$\pm$0.064&4.206 ($\theta_{23}^\textrm{CKM}=$2.401$^\circ$)&4.204 ($\theta_{23}^\textrm{CKM}=$2.408$^\circ$)\\ 
\rowcolor{yellow!13}$\theta_{13}^\textrm{CKM}/10^{-3}$&3.64$\pm$0.13&3.64 ($\theta_{13}^\textrm{CKM}=$0.208$^\circ$)&3.64 ($\theta_{13}^\textrm{CKM}=$0.208$^\circ$)\\ 
\rowcolor{yellow!13}$\delta_\textrm{CKM}$&1.208$\pm$0.054&1.209 ($\delta_\textrm{CKM}=$69.322$^\circ$)&1.212 ($\delta_\textrm{CKM}=$69.457$^\circ$)\\ 
 \hline\hline

\rowcolor{green!10}$\Delta m^2_{21}/10^{-5} $&7.425$\pm$0.205&7.413&7.506\\ 
\rowcolor{green!10}$\Delta m^2_{31}/10^{-3}$ (NO)&2.515$\pm$0.028&2.514&-\\
\rowcolor{green!10}$\Delta m^2_{32}/10^{-3}$ (IO)&-2.498$\pm$0.028&-&-2.499\\ 
\hline\hline

\rowcolor{orange!13}$\sin^2 \theta_{12}$ &0.3045$\pm$0.0125&0.3041 ($\theta_{12}=33.46^\circ$)&0.3067 ($\theta_{12}= 33.63^\circ$)\\ 
\rowcolor{orange!13}$\sin^2 \theta_{23}$ (NO)$^*$  &0.5705$\pm$0.0205&0.4473 ($\theta_{23}=41.98^\circ$)&-\\ 
\rowcolor{orange!13}$\sin^2 \theta_{23}$ (IO)$^*$ &0.576$\pm$0.019&-&0.5784 ($\theta_{23}= 49.51^\circ$)\\
\rowcolor{orange!13}$\sin^2 \theta_{13}$ (NO)&0.02223$\pm$0.00065&0.02223 ($\theta_{13}=8.57^\circ$)&-\\  
\rowcolor{orange!13}$\sin^2 \theta_{13}$ (IO)&0.02239$\pm$0.00063&-&0.02238 ($\theta_{13}= 8.60^\circ$)\\

\rowcolor{orange!13}$\delta^\circ_\mathrm{CP}$ (NO)&207.5$\pm$38.5&240.49&-\\
\rowcolor{orange!13}$\delta^\circ_\mathrm{CP}$ (IO)&284.5$\pm$29.5&-&263.49\\

\hline\hline
\rowcolor{cyan!40}$\eta_B/10^{-10}$&6.12$\pm$0.04$^\ddagger$&7.6 (7.6)& 9.6 (51)\\

\hline \hline
\rowcolor{white!10}$\chi^2$&-&1.45&5.76$^\dagger$ \\
\hline

\end{tabular}
}
\caption{Fitted values of the observables.  See text for details. ${}^*$Note that experimental measurements of $\theta_{23}$ have two local minima~\cite{NUFIT}, one is for smaller and the other for larger values than $45^\circ$. For these experimental values, although only the best fit values from the global fit~\cite{NUFIT} are shown in this table, in the fitting procedure we have allowed the entire viable ranges.    ${}^\dagger$For the inverted neutrino mass ordering, a significant contribution to  $\chi^2$, in particular, $\left(\Delta\chi^2\right)_{\theta_{23}}\approx 2.7$, originates from global fit to neutrino parameters~\cite{NUFIT}. $^\ddagger$Since the uncertainty of the observed baryon asymmetry parameter $\eta_B$ from Planck~\cite{Planck:2018vyg} is much smaller than the expected ``theoretical uncertainty'' alluded to in the previous section, we have not added its contribution to the total $\chi^2$ quoted above (see text for more details). 
The generated $\eta_B$ values quoted here are for zero (thermal) initial $N_i$ abundance.  
}\label{result}
\end{table}

We then perform a minimization of a $\chi^2$-function at the $M_Z=91.8176$ GeV scale, which is defined as 
\begin{align}
\chi^2= \sum_k \left( \frac{T_k-O_k}{E_k} \right)^2,    
\end{align}
where $T_k$, $O_k$, and $E_k$ stand for theoretical prediction, experimentally observed central value, and $1\sigma$ experimental uncertainty, respectively,  for the $k$-th physical quantity.  The sum over $k$ includes 3 up-type quark, 3 down-type quark, 3 charged lepton masses, 3 CKM (Cabibbo–Kobayashi–Maskawa) mixing angles, 1 Dirac phase in the CKM matrix, 2 neutrino mass-squared differences, 3 PMNS mixing angles, and 1 Dirac phase in the  PMNS matrix.  Low scale experimental values of these observables in the charged and neutral fermion sectors are taken from Refs.~\cite{Antusch:2013jca} and~\cite{NUFIT,Esteban:2020cvm}, respectively. The $\chi^2$-function also includes the baryon asymmetry of the universe, $\eta_B$. Since the asymmetry is generated from the decay of the next-to-lightest right-handed neutrinos, $\eta_B$ is computed using the Dirac Yukawa coupling $Y_{\nu_D}$ and the right-handed neutrino masses $M_{\nu_R}$ evaluated  at $M_2$, the mass scale of $N_2$.

As pointed out first in Ref.~\cite{Saad:2022mzu}, this minimal Yukawa sector of $SO(10)$ allows solutions for both normal ordering (NO) and inverted ordering (IO) of the light neutrino masses. In this work, we also consider these two scenarios and benchmark fit results  are given in Table~\ref{result}. This table also includes the experimentally measured values of the observables and their associated $1\sigma$ uncertainties. Since the experiment errors on the charged fermion masses are very small, we have quoted $1\%$ uncertainty in their masses in the table,  which we have used. 
The fit parameters for these two cases  (NO and IO) are provided in Appendices~\ref{A} and~\ref{B}, respectively.

The right-handed neutrino mass spectra obtained from the best fits are
\begin{align}
\mathrm{NO}:& \left(M_1, M_2, M_3\right)= \left( 6.57\times 10^4, 2.08\times 10^{12}, 8.10\times 10^{14} \right)\; \mathrm{GeV}, 
\\
\mathrm{IO}:& \left(M_1, M_2, M_3\right)= \left( 1.06\times 10^4, 1.72\times 10^{12}, 5.85\times 10^{14} \right) \; \mathrm{GeV}.
\end{align}
The light neutrino masses $m_i$, the effective neutrino mass appearing in neutrinoless double beta decay, $m_{\beta\beta}$, and the Majorana phases $\alpha$ and $\beta$ (cf. Eq. (\ref{Upmns-convention})) are given as follows:
\begin{align}
\mathrm{NO}:& \left(m_1, m_2, m_3, m_{\beta\beta}\right)= \left(0.038, 8.61, 50.14, 3.68\right) \; \mathrm{meV}, \; (\alpha,\beta)=(178.55^\circ, 124.46^\circ),\label{eq:predictions_NO}
\\
\mathrm{IO}:& \left(m_1, m_2, m_3, m_{\beta\beta} \right)= \left(49.24, 49.99, 0.1923, 34.36\right) \; \mathrm{meV}, \; (\alpha,\beta)=(100.76^\circ, 172.86^\circ).\label{eq:predictions_IO}
\end{align}

\begin{figure}[t!]
\centering
\includegraphics[width=12cm]{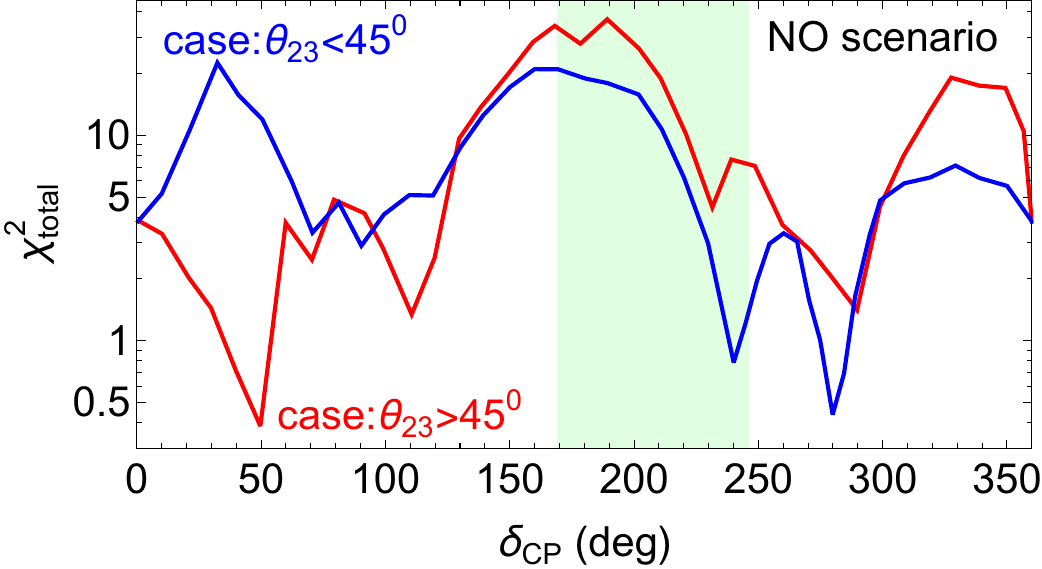}
\caption{  Variation of the Dirac CP-violating phase, $\delta_\mathrm{CP}$, in the neutrino sector obtained by
marginalizing over all other model parameters for the case of NO solution. The $1\sigma$ range of $\delta_\mathrm{CP}$ from global fit is depicted with light green band~\cite{NUFIT}.  For making this plot, in the numerical fit procedure, we have utilized analytical approximated formula for $\eta_B$, and demanded that the baryon asymmetry lies within the range $\eta_B/10^{-10}\in (5,50)$. One should not get confused by the fact that at $\delta_\mathrm{CP}\sim 280^\circ$, one sees a deeper minimum compared to the solution presented in Table~\ref{result}. This is simply because the benchmark fit presented in Table~\ref{result} is closer to the central value of $\delta_\mathrm{CP}$ obtained from global fit of neutrino observables. The blue (red) curve represents the solutions corresponding to the case $\theta_{23}< 45^\circ$ ($\theta_{23}> 45^\circ$).  }
\label{fig:phase}
\end{figure}

Concerning the solutions presented in Table~\ref{result}, note that  the total $\chi^2$ per degrees of freedom for both the NO and IO solutions are similar.  While the model admits both NO and IO spectra, global fits  and cosmological constraints show a preference of NO over IO. A variation of the Dirac CP-violating phase, $\delta_\mathrm{CP}$, in the neutrino sector obtained for the case of normal ordering by
marginalizing over all other model parameters is shown in Fig.~\ref{fig:phase}. This plot shows a slight preference of the phase close to  the region $\delta_\mathrm{CP}\in (230-300)^\circ$ for the solutions corresponding to $\theta_{23}< 45^\circ$ (the blue curve in Fig.~\ref{fig:phase}). Although for NO, solutions are also possible for the case of  $\theta_{23}> 45^\circ$ (the red curve in Fig.~\ref{fig:phase}), comparatively it has  somewhat larger total $\chi^2$ within the $1\sigma$ preferred range of $\delta_\mathrm{CP}$ from global fit (light green band in Fig.~\ref{fig:phase}).   In this particular case, the preferred value of the CP violating phase is rather close to $\delta_\mathrm{CP}\sim 50^\circ$.

For leptogenesis, the washout parameters defined in Eq.~\eqref{eq:washout_parameter} are  obtained from the best fit and are given by
\begin{eqnarray}
    \mathrm{NO}: K_1(\mu = M_2) \simeq 19, \quad
    K_2(\mu = M_2) \simeq 0.39, \quad
    K_3(\mu = M_2) \simeq 63, \\
    \mathrm{IO}: K_1(\mu = M_2) \simeq 70, \quad
    K_2(\mu = M_2) \simeq 0.65, \quad
    K_3(\mu = M_2) \simeq 88.
\end{eqnarray}
At $\mu = M_1$, while $K_2$ and $K_3$ remain the same, we have $K_1(\mu = M_1) \simeq 18$ and $K_1(\mu = M_1) \simeq 64$ for NO and IO, respectively. 
Notice that we found $M_3 \simeq 8  \times 10^{14}\,{\rm GeV}$ and $5  \times 10^{14}\,{\rm GeV}$ for NO and IO, respectively. 
One could wonder whether for a sufficiently large reheat temperature,
comparable to $M_3$, one should also include a contribution to the $B-L$ asymmetry from $N_3$-decays. However, 
we have checked that the $B-L$ asymmetry generated from $N_3$-decays is in any case negligible. This is because for $M_3 \gg M_2$ 
all $N_3$ $C\!P$ flavored asymmetries are  strongly suppressed. As for $N_1$, since $M_1 \ll 10^9$ GeV, the CP asymmetries 
are too small for leptogenesis, as one could expect from the usual  Davidson-Ibarra lower bound 
on right-handed neutrino masses for successful leptogenesis \cite{Davidson:2002qv}. Notice that since $K_1 \gg 1$, 
flavor effects are crucial to avoid the $N_1$ wash-out, 
since the wash-out suppression is exponential in $K_1$ \cite{DiBari:2005st,Vives:2005ra}. 
We will discuss this point in more detail in Sec.~\ref{sec:SO10inspired}.
Finally, notice that since  $M_2 \sim 10^{12}$ GeV and thanks to the interference with $N_3$, 
the $N_2$ decays can generate the correct asymmetry for successful leptogenesis. 
Since $K_2$ is close to the boundary between strong and weak washout regimes, one  expects some sensitivity to the initial $N_2$ abundance. We indeed observe an order of magnitude enhancement in the asymmetry generation for the case of thermal  initial $N_2$ abundance compared to the case of initial vanishing 
$N_2$ abundance at the production. This is because in this case the small 
neutrino Yukawa interactions are not responsible for the production of the $N_2$ abundance that is assumed to 
be thermal prior to $N_2$ decays. 
This should not be regarded as a drawback  of the scenario since the thermalisation of the $N_2$
abundance can be achieved thanks to the $SO(10)$ gauge interactions, as we previously discussed, if the 
reheat temperature is sufficiently large (see Eq.~(\ref{Ntherm})). 

Solving the Boltzmann equations discussed in Sec.~\ref{sec:lepto}, the resulting values of baryon to photon number density today assuming zero (thermal) initial $N_i$ abundances are found for the best fit to be
\begin{eqnarray}
    \mathrm{NO}: \eta_B = 7.6\, (7.6)
    \times 10^{-10}, \qquad
    \mathrm{IO}: \eta_B = 9.6\, (51) \times 10^{-10},
\end{eqnarray}
which are slightly larger than the observed value $(6.12 \pm 0.04) \times 10^{-10}$~\cite{Planck:2018vyg}. 
Due to the ``theoretical uncertainty'' mentioned 
at the end of the previous section, we have not included this measurement into the fit. 
Instead, we just require the generated asymmetry to be of the correct sign and not smaller than the observed value.\footnote{While a larger baryon asymmetry can always be reduced by additional entropy injection and/or washout processes after baryogenesis, a scenario which generates an insufficient asymmetry is a failed baryogenesis by definition.}
Though we did observe enhanced asymmetry generation from $N_2$ leptogenesis with thermal initial $N_2$ abundance for both NO and IO cases, curiously, nontrivial flavor effects during $N_1$ washout renders the final asymmetry to be the same as that of zero initial $N_2$ abundance in the NO case while some enhancement survives for the IO case. 
What we observe in the \emph{charged lepton flavor basis} where $y_E$ is diagonal is the following. The $N_1$ washout of the initially dominant $(Y_{\tilde \Delta} - Y_{\Delta E})_{\tau\tau}$ (from $N_2$ leptogenesis) is strong for both NO and IO scenarios, which makes it eventually subdominant. 
On the other hand, the $N_1$ washout of $(Y_{\tilde \Delta} - Y_{\Delta E})_{\mu\mu}$ is under control for both NO and IO scenarios, eventually making it the dominant one with positive sign.
Finally, the $N_1$ washout of $(Y_{\tilde \Delta} - Y_{\Delta E})_{ee}$ is weak and strong in the NO and IO scenarios, respectively, making it relevant only for the NO scenario. Due to the negative sign in the final $(Y_{\tilde \Delta} - Y_{\Delta E})_{ee}$, it partially cancels $(Y_{\tilde \Delta} - Y_{\Delta E})_{\mu\mu}$ in the NO scenario such that the final asymmetry in both zero and thermal $N_i$ remains similar.
This is another example of the importance of flavor effects in leptogenesis.
In Sec.~\ref{sec:SO10inspired} we will provide some analytical insight on these results on the final asymmetry, showing how flavor effects can explain it.

\subsection{Features of the minimal Yukawa sector}
In the following, we illustrate two unique  features of the minimal Yukawa sector, namely (i) the right-handed neutrino spectrum is predicted  and (ii) contrary to other $SO(10)$ setups, inverted neutrino mass ordering solution is possible within this framework. First, one important feature of this model is that, a viable fit to fermion mass spectrum can obtained provided that $2\times 10^{13}\;\mathrm{GeV} \lesssim M_3 \lesssim 10^{15}\;\mathrm{GeV}$~\cite{Saad:2022mzu}, while best fit typically prefers $M_3\sim 10^{14}$ GeV. Moreover, for the moment, focusing only on orders of the diagonal entries (hence antisymmetric contribution is ignored), the up-type quark, down-type quark, and charged lepton mass matrices are,
\begin{align}
&M_U\sim D+S,\;\;\;\;\;
M_D\sim D+r_1 S,\;\;\;\;\;
M_E\sim D-3 r_1 S.
\end{align}
It is obvious from these relations that the only consistent solution can be found if $D\ll S$ and $r_1\ll 1$. One, therefore, has 
$D_{33}+r_1S_{33}\sim m_b$ (and similarly, $D_{33}-3r_1S_{33}\sim m_\tau$), whereas $S_{33}\sim m_t$ (for the top quark mass,  $D_{33}$ plays no role owing to its smallness). Due to the smallness of strange-quark mass compared to the charm-quark mass, a similar argument is also valid for the second generation (however, such an argument is not applicable to the first generation). Therefore, reproducing the correct fermion mass hierarchies demands, 
\begin{align}
S\sim \begin{pmatrix}
S_{11}&&\\
&m_c&\\
&&m_t
\end{pmatrix}.    
\end{align} 
Then the ratio $r_1$ must obey $r_1\lesssim 10^{-2}$ not to provide too large contributions to the  masses in the down-type quark sector. An immediate consequent is that 
\begin{align}
M_2\sim \frac{m_c}{m_t} M_3\sim 10^{11-12}\;\mathrm{GeV},     
\end{align}
which is a prediction of this scenario. The features mentioned above are true for both NO and IO solutions.

Furthermore, the $S_{11}$ entry plays no role in charged fermion mass fit, which is solely fixed by the requirement of reproducing correct neutrino fit. Looking at the light neutrino mass matrix at the GUT scale and using the fit parameters, we obtain (for illustration, we only present the order of the numbers)
\begin{align}
&m_\nu= - M^T_{\nu_D} \begin{pmatrix}
M^{-1}_1&&\\  
&M^{-1}_2&\\
&&M^{-1}_3
\end{pmatrix} M_{\nu_D},
\\&\sim  
\left(
\begin{array}{ccc}
 \frac{3\times 10^{-7}}{M_1}\color{gray}+10^{-17} & \frac{4\times 10^{-7}}{M_1}\color{gray}+10^{-15} & \frac{4\times 10^{-7}}{M_1}\color{gray}+10^{-15} \\
  & \frac{4\times 10^{-7}}{M_1}\color{gray}+10^{-13} & \frac{6\times 10^{-7}}{M_1}\color{gray}+10^{-13} \\
 &  & \frac{6\times 10^{-7}}{M_1}+10^{-10} \\
\end{array}
\right)_\mathrm{NO}, 
\\&\xrightarrow[]{M_1^\mathrm{NO} \sim 10^5\;\mathrm{GeV}}
\left(
\begin{array}{ccc}
 5\times 10^{-12} &6\times 10^{-12} &6\times 10^{-12} \\
  & 9\times 10^{-12} & 9\times10^{-12} \\
 &  &8\times 10^{-11} \\
\end{array}
\right)_\mathrm{NO}. 
\end{align}
The contributions from $M_{2,3}$ are presented in gray. It turns out that apart from the (3,3) entry, contributions from $M_{3}$ are completely negligible. However, subleading contributions from $M_2$ on the (2,2), and (2,3) entries cannot be fully ignored. For the rest of the terms, $M_1$ contributions dominate over the rest. Hence, the mass of the lightest right-handed neutrino, $M_1$, plays a vital role in generating neutrino masses via type-I seesaw formula. Therefore, $M_1$ cannot be larger than a certain value. To provide sufficient contributions and reproduce neutrino oscillation data, for the NO solution, one must have $M^\mathrm{NO}_1\sim 10^{5}$ GeV. This accordingly fixes $S_{11}\sim 10^{-9}$, making it irrelevant for charged fermion masses. 

On the other hand, for IO case, $M^\mathrm{IO}_1$ must be about an order of magnitude smaller than $M^\mathrm{NO}_1$ such that it provides dominant contributions in the first row and first column of $m_\nu$. In particular, using the fit parameters, we find (as before, we only present the order of the numbers)
\begin{align}
&m_\nu\sim  
\left(
\begin{array}{ccc}
 \frac{5\times 10^{-7}}{M_1}\color{gray}+10^{-17} & \frac{7\times 10^{-8}}{M_1}\color{gray}+10^{-15} & \frac{4\times 10^{-7}}{M_1}\color{gray}+10^{-15} \\
  & \frac{10^{-8}}{M_1}\color{gray}+10^{-13} & \frac{6\times 10^{-8}}{M_1}\color{gray}+10^{-13} \\
 &  & \frac{3\times 10^{-7}}{M_1}+10^{-10} \\
\end{array}
\right)_\mathrm{IO}, 
\\&\xrightarrow[]{M_1^\mathrm{IO}\sim 10^4\;\mathrm{GeV}}
\left(
\begin{array}{ccc}
 5\times 10^{-11} & 8\times 10^{-12} &4\times 10^{-11} \\
  & 10^{-12} & 6\times 10^{-12} \\
 &  & 3\times 10^{-11} \\
\end{array}
\right)_\mathrm{IO}. 
\end{align}
This shows that the model predicts $M_{1}\sim 10^5$ GeV for NO and $M_{1}\sim 10^4$ GeV for IO. Summarizing, the model predicts,
\begin{align}
\left(M_1, M_2, M_3\right)\sim     \left(10^{4-5}, 10^{11-12}, 10^{14-15}\right)\;\textrm{GeV}. 
\end{align}

It is intriguing to point out that such a high value $\sim 10^{15}$ GeV of the $B-L$ symmetry breaking scale could explain the recently observed gravitational waves at nanoHertz frequencies (see, for example, Ref.~\cite{Antusch:2023zjk}).  A series of pulsar timing arrays, CPTA~\cite{Xu:2023wog}, EPTA~\cite{EPTA:2023fyk}, NANOGrav~\cite{NANOGrav:2023gor}, and PPTA~\cite{Reardon:2023gzh}, very recently reported this result, which could originate from metastable cosmic strings with a string tension, $\mu_\mathrm{cs}$, such that $G\mu_\mathrm{cs}\sim 10^{-7}$~\cite{NANOGrav:2023hvm} (here, $G$ is the Newton's constant, and, $G\mu_\mathrm{cs}$ is the dimensionless string tension parameter), corresponding to a metastable cosmic string network formation scale of $\sim 10^{15}$ GeV.

\section{Comparison with $SO(10)$-inspired leptogenesis}\label{sec:SO10inspired}
In this section we compare the obtained numerical results with the analytical results obtained within
the framework of $SO(10)$-inspired leptogenesis \cite{DiBari:2014eya,DiBari:2017uka,DiBari:2020plh}.
First of all, it is convenient to define the quantities $N_X \equiv Y_X/Y^{\rm eq}_{N_2}\, (T\gg M_2)$, so that
simply $N_{N_2}^{\rm eq}(T\gg M_2) = 1$.  The final $B-L$ asymmetry has to be calculated
as the sum of the three charged lepton flavor asymmetries 
\begin{equation}\label{NBmLf}
N_{B-L}^{\rm f} = \sum_{\alpha = e,\mu, \tau} \, N_{\Delta_\alpha} \,  ,
\end{equation}
where $\Delta_{\alpha} \equiv B/3 - L_{\alpha}$.  The  baryon-to-photon ratio predicted by leptogenesis 
can then be calculated as  $0.96\times 10^{-2}\,N_{B-L}^{\rm f}$.  We also define the neutrino Dirac mass matrix in the charged lepton mass basis as $m_D \equiv (M_{\nu_D}E_L)^\dagger$ 
and define the corresponding light neutrino mass matrix in the flavor basis  as
$\bar m_\nu \equiv - m_D M^{-1} m_D^T$ in a way that, in terms of $\bar m_\nu$,
the light neutrino Majorana mass term is $ \overline{\nu}\, \bar m_{\nu}\,\nu^c /2$ with mass eigenvalues $D_m \equiv \mathrm{diag}(m_1,m_2,m_3)$. 
Within $SO(10)$-inspired leptogenesis  it is assumed 
that if one diagonalises the neutrino Dirac mass matrix with a biunitary 
transformation, $m_D =$\\ $V_L^\dagger {\rm diag}(m_{D1}, m_{D2}, m_{D3})\,U_R$, then 
$(m_{D1}, m_{D2}, m_{D3}) \simeq (m_{u}, m_{ c}, m_{t})$ and 
$V_L \simeq U_{\rm CKM}$ \cite{Buchmuller:1996pa,Nezri:2000pb,Buccella:2001tq,Branco:2002kt,Akhmedov:2003dg,DiBari:2008mp}. 

The spectrum of right-handed neutrino masses resulting from the seesaw formula in combination with
the low energy neutrino experimental data is very hierarchical with $M_1 \ll 10^9 \,{\rm GeV}$
and $M_2 \sim 10^{11}$--$10^{12}\,{\rm GeV}$, exactly of the kind obtained in the numerical fit. 
With such a hierarchical spectrum, the neutrino Dirac mass eigenvalue that enters the final
asymmetry is $m_{D2} \sim m_{c}$, while $m_{D1}$ and $m_{D3}$ cancel out and do not play any role: for this
reason there is a reduction of the number of parameters that determines the asymmetry and once
the condition of successful leptogenesis is imposed, there are some interesting constraints on low energy neutrino parameters.  
From this point of view the numerical fit that we have obtained seems to respect the main features
of the solutions obtained in $SO(10)$-inspired leptogenesis.  In particular, NO solutions are favoured
and first octant for $\theta_{23}$ is also favoured.  
 There is, however, one remarkable departure from the solutions obtained within  $SO(10)$-inspired leptogenesis.
 The final asymmetry is either tauon dominated, with $N_{B-L}^{\rm f} \simeq  N_{\Delta_\tau}$,
 or muon dominated, with  $N_{B-L}^{\rm f} \simeq  N_{\Delta_\mu}$.  In both cases one obtains
 a lower bound on the absolute neutrino mass scale that in terms of lightest neutrino
 mass can be expressed as $m_1 \gtrsim 1\,{\rm meV}$ 
 in the case of tauon-dominated solutions and $m_1 \gtrsim 20\,{\rm meV}$ in the case of muon-dominated solutions,
 where we refer to the, by far more interesting, NO case.

In the numerical fit we have obtained, the asymmetry is muon-dominated but as one can see one has 
$m_1 \simeq 0.038\,{\rm meV}$, violating the lower bound obtained for $SO(10)$-inspired leptogenesis 
with $V_L \simeq U_{\rm CKM}$.  If we parameterise the unitary matrix $V_L$ in terms of three mixing angles
and six phases, we can write:
\begin{eqnarray}\label{VLmatrix}
V_L&=&
\left( \begin{array}{ccc}
c^L_{12}\,c^L_{13} & s^L_{12}\,c^L_{13} & s^L_{13}\,e^{-{\rm i}\,\delta_L} \\
-s^L_{12}\,c^L_{23}-c^L_{12}\,s^L_{23}\,s^L_{13}\,e^{{\rm i}\,\delta_L} &
c^L_{12}\,c^L_{23}-s^L_{12}\,s^L_{23}\,s^L_{13}\,e^{{\rm i}\,\delta_L} & s^L_{23}\,c^L_{13} \\
s^L_{12}\,s^L_{23}-c^L_{12}\,c^L_{23}\,s^L_{13}\,e^{{\rm i}\,\delta_L}
& -c^L_{12}\,s^L_{23}-s^L_{12}\,c^L_{23}\,s^L_{13}\,e^{{\rm i}\,\delta_L}  &
c^L_{23}\,c^L_{13}
\end{array}\right)
\nonumber \\ 
&& \times {\rm diag}\left(e^{-i\,{\alpha_L\over 2}}, e^{-i\,{\beta_L\over 2}}, 1  \right)\,   ,
\end{eqnarray}
where $c_{ij}^L = \cos \theta^L_{ij}$ and $s_{ij}^L = \sin\theta^L_{ij}$.
We have extracted this matrix in the case of the best numerical fit and found that 
while  $\theta_{12}^L \simeq 4^\circ$ and $\theta_{13}^L \simeq 0.3^\circ$ respect
the condition to be comparable to the corresponding values of the angles in the CKM matrix, 
one has $\theta_{23}^L \simeq 45^\circ$, a maximal value corresponding to a drastic departure
from the condition $V_L \simeq U_{\rm CKM}$. It was already noticed that a non-vanishing 
value of $\theta_{23}^L$ is necessary to have non-vanishing $\varepsilon_{2\mu}$ and, therefore,
muon dominated solutions \cite{DiBari:2017uka}. However, this angle was still kept small and it was
shown one could not obtain successful leptogenesis in the 
normal hierarchical limit ($m_1 \ll m_{\rm sol} \equiv \sqrt{\Delta m^2_{21}} \sim 10\,{\rm meV}$).
Here we want to show how, thanks to a large value of $\theta_{23}^L$, one can have successful
(muon dominated) leptogenesis with $m_1$ as small as $0.038\,{\rm meV}$, as obtained in the numerical fit. 

If in first approximation we neglect flavor coupling stemming from spectator processes, mainly dominated
by the Higgs asymmetry, an analytical expression for the muonic contribution is given by
\begin{equation}
N_{\Delta_\mu}^{\rm lep,f} \simeq   \varepsilon_{2\mu}\,\kappa(K_{2e} + K_{2\mu}) 
\, e^{-{3\pi\over 8}\,K_{1 \mu}} \,  ,
\end{equation}
where $K_{1\mu}$, $K_{2\mu}$ and $K_{2e}$ are the flavored decay parameters
given by $K_{i\alpha} = |(m_{D})_{\alpha i}|^2/(M_i\,m_\star)$,
$\varepsilon_{2\mu}$ is the muonic $C\!P$ asymmetry and $\kappa(K_{2e} + K_{2\mu})$
is the efficiency factor at the production, i.e., at the scale $M_2$. 
For initial thermal $N_2$ abundance this can be calculated as
\be\label{kappa}
\k(K_{2\a}) = 
{2\over z_B(K_{2e} + K_{2\mu})\,(K_{2e} + K_{2\mu})}
\left(1-e^{-{(K_{2e} + K_{2\mu})\,z_B(K_{2e} + K_{2\mu})\over 2}}\right) \,  , 
\ee
with
\be
z_B(K_{2e} + K_{2\mu}) \simeq 
2+4\,(K_{2e} + K_{2\mu})^{0.13}\,e^{-{2.5\over K_{2e} + K_{2\mu}}} \,   .
\ee
We have now to understand the effect of a maximal $\theta_{23}^L = 45^\circ$. 
For simplicity we can approximate $\theta_{12}^L \simeq \theta_{13}^L \simeq 0$. 
A very useful quantity in connecting the neutrino Dirac mass matrix  to the low energy neutrino parameters
via the seesaw formula when $V_L \neq I$ is the neutrino mass matrix in the Yukawa basis
given by  $\widetilde{m}_{\nu} = V_L \, \bar m_{\nu} \, V_L^T$, that is clearly symmetric. 
In particular,  the right-handed neutrino masses can be expressed in a compact way in terms of the $\widetilde{m}_{\nu}$ entries as:
\be\label{Mi}
M_1    \simeq   {m^2_{D1} \over |(\widetilde{m}_{\nu})_{11}|} \, , \;\;
M_2  \simeq    {m^2_{D2} \over m_1 \, m_2 \, m_3 } \, 
{|(\widetilde{m}_{\nu})_{11}| \over |(\widetilde{m}_{\nu}^{-1})_{33}|  } \,  ,  \;\;
M_3  \simeq   m^2_{D3}\,|(\widetilde{m}_{\nu}^{-1})_{33}|  \,  .
\ee
Since we take only $\theta_{23}^L$ large, it is easy to see 
that the only entries  $(\widetilde{m}_{\nu})_{ij}$  that change compared to the case $V_L = I$ are
$(\widetilde{m}_{\nu})_{23}$ and $(\widetilde{m}_{\nu})_{33}$. With this observation it is simple to see the effect of 
having a large $\theta_{23}^L$ in the muon-dominated final asymmetry.

Let us start from calculating $K_{1\mu}$. It is simple to derive the following 
approximate expression in the hierarchical limit ($m_1 \ll m_{\rm sol}$) \cite{DiBari:2017uka}:
\begin{equation} 
K_{1\mu} = {|(m_{D})_{\mu 1}|^2 \over M_1\,m_\star}  \simeq c^2_{12}\,c^2_{23}\,{m_{\rm sol}\over m_{\star}} \simeq 3 \, .
\end{equation}
In this expression we have neglected a correcting term $ (s^{L}_{23})^2 \, |(\widetilde{m}_{\nu})_{13}|^2/(|(\widetilde{m}_{\nu})_{11}|m_\star)$, 
so that there is no substantial change in $K_{1\mu}$ even with a maximal value $\theta_{23}^{L} = 45^\circ$.  Notice that this result implies that there is necessarily a suppression of the muon asymmetry produced from $N_2$ decays due to the $N_1$ washout 
given by a factor $e^{-3\pi\,K_{1\mu}/8} \simeq 0.03$. This suppression of course needs to be compensated mainly by a 
large enhancement of the muonic $C\!P$ asymmetry $\varepsilon_{2\mu}$ and, to a minor extent, by a small value of 
$K_{2e} + K_{2\mu}$ so that $\kappa(K_{2e} + K_{2\mu}) \simeq 1$ for initial thermal $N_2$ abundance.

Let us show that this is indeed what happens starting from the crucial quantity $\varepsilon_{2\mu}$. 
An approximate expression for $\varepsilon_{2\mu}$ is given by
\be
\varepsilon_{2\mu} \simeq {3\over 16\,\pi}\,{M_2\over M_3} \,
 {{\rm Im}\left[(m_D)_{\a 2}^{\star}
(m_D)_{\a 3}(m_D^{\dag}\, m_D)_{23}\right]\over v^2\, (m_D^{\dag}\, m_D)_{22} }\,   .
\ee
From this expression, using the Eqs.~(\ref{Mi}) for $M_2$ and $M_3$, the biunitary parameterisation for $m_{\rm D}$
and an expression of $U_R$ in terms of $\widetilde{m}_{\nu}$ entries, one arrives to the following approximate expression
for $\ve_{2\mu}$:
\be
\ve_{2\mu} \simeq    {3 \,m^2_{D2} \over 16\, \pi \, v^2}\,
{|(\widetilde{m}_{\nu})_{11}| \, |({V_L})_{32}|^2 \over m_1 \, m_2 \, m_3 \, |(\widetilde{m}_{\nu}^{-1})_{33}|^2}  \sin \alpha_{\rm lep}\,   ,
\ee 
where  from Eq.~(\ref{VLmatrix}) one can see that, in our case, $|({V_L})_{32}|^2 = (s^L_{23})^2 \sim 1$. 
Moreover, one finds for the effective phase 
\be
\a_{\rm lep}  = 
{\rm Arg}\left[(\widetilde{m}_{\nu})_{11} \right]  - 2\,{\rm Arg}[(\widetilde{m}^{-1}_{\nu})_{23}] 
- \pi + \alpha -2\,\beta  + \alpha_L - 2\,\beta_L  \,   .
\ee
It should be immediately noticed how the $C\!P$ asymmetry vanishes in the limit $\theta_{23}^L \ra 0$ showing how
muon-dominated solutions necessarily require some departure from the case $V_L = I$. We now have to understand
how one can have an enhancement in the hierarchical limit, for $m_1 \ll m_{\rm sol}$.
The crucial point is to have a small value of $|(\widetilde{m}_{\nu}^{-1})_{33}|$. Let us calculate this important quantity.
First of all notice that $\widetilde{m}_{\nu}^{-1} = V^\star_L\, \bar m_\nu^{-1} \, V_L^\dagger$ so that one obtains 
\be
(\widetilde{m}_{\nu}^{-1})_{33} = (V^\star_{L})_{33}^2 \, (\bar m_{\nu}^{-1})_{33} + (V^\star_{L})_{32}^2\, (\bar m_{\nu}^{-1})_{22} 
+ 2\,(V^\star_{L})_{32}(V^\star_{L})_{33} \, (\bar m_{\nu}^{-1})_{23} \,  . 
\ee
The inverse low energy neutrino mass matrix is given by $\bar m_{\nu}^{-1} = U_\nu^\star \, D_{\rm m}^{-1} \, U_\nu^\dagger$, so that one finds
in the hierarchical limit:
\be
(\bar m_{\nu}^{-1})_{33} \simeq  
{s^2_{12}s^2_{23}\over m_1}\,e^{i\a} \,  , \;\;
(\bar m_{\nu}^{-1})_{22} \simeq 
{s^2_{12}\,c^2_{23}\over m_1}\,e^{i\a} \,   ,  \;\; 
(\bar m_{\nu}^{-1})_{23} \simeq  
-{s^2_{12}\,s_{23}\,c_{23} \over m_1}\,e^{i\a}  \,  .
\ee
Notice that for $m_1 \ra 0$, one has  $(\widetilde{m}_{\nu}^{-1})_{33} \propto 1/m_1$ implying
$\ve_{2\mu} \propto m_1 \ra 0$. It is then necessary that the factor of proportionality 
$(\widetilde{m}_{\nu}^{-1})_{33} \, m_1 $ gets sufficiently small in order to have successful leptogenesis. 
This also corresponds to have a reduction of the ratio $M_2 / M_3 \propto (\widetilde{m}_{\nu}^{-1})_{33}^2$.
This implies
that there must be some phase cancellation. Let us see whether this is possible for some choice of the values of some of the phases.

In the hierarchical limit one obtains then the following expression for $(\widetilde{m}_{\nu}^{-1})_{33}$:
\be
(\widetilde{m}_{\nu}^{-1})_{33}\, m_1 \simeq s^2_{12}\,e^{i\a}\left[(c^L_{23})^2 \, s^2_{23} 
+ 2 s^L_{23}\,c^L_{23} s_{23} c_{23}\,e^{i{\beta_L\over 2}} + (s^L_{23})^2 \, c^2_{23} \,e^{i{\beta_L}}  \right] \,  .
\ee
For $\theta_{23}^L = \pi/4$, one has a cancellation in the second term if $\beta_L \simeq 2n\,\pi$ with $n$ integer number in a way that one obtains
\be
|(\widetilde{m}_{\nu}^{-1})_{33}|\, m_1 \simeq {s^2_{12}\over 2} (c_{23}-s_{23})^2 \sim 0.01 \,  ,
\ee
where for the numerical estimation we used the best fit values for $\theta_{12}$ and $\theta_{23}$. 
Notice that in our fit we indeed find $\beta_L \simeq -2\pi$ and $|(\widetilde{m}_{\nu}^{-1})_{33}|\, m_1 \sim 0.01$, in nice agreement with this analytical result. 
 The smallness of $|(\widetilde{m}_{\nu}^{-1})_{33}|\, m_1$, combined with a large $\theta_{23}^L \simeq \pi/4$,
 strongly enhances the muonic $C\!P$ asymmetry, allowing successful leptogenesis despite that for 
 $m_1 \ll m_{\rm sol}$ one has $K_{1\m} \simeq 3$, as we have seen. Notice that a small $|(\widetilde{m}_{\nu}^{-1})_{33}|$ is also important to suppress $K_{2e}+K_{2\mu}$ since one has in the hierarchical limit
 \be
K_{2\mu} = {|(m_{D})_{\mu 2}|^2 \over M_2\,m_\star}  \simeq (c^L_{23})^2\, {m_1\,m_{\rm sol}\,m_{\rm atm} \over m_{\star}\,|(\widetilde{m}_{\nu})_{11}|}\,|(\widetilde{m}_{\nu}^{-1})_{33}| \sim 0.1 \,  ,
 \ee
 and $K_{2e} \ll K_{2\mu}$. It is interesting that the value $\theta_{23}^L$ seems to be a special one in order to have such phase cancellation
 and this might suggest the presence of some discrete symmetry. It is also interesting that in this case the deviation of the atmospheric mixing angle
 from the maximal value sets the value of $|(\widetilde{m}_{\nu}^{-1})_{33}|\, m_1$ and, consequently, of the baryon asymmetry.\footnote{Notice that
 for $\theta_{23} =\pi/4$ one would have an exact cancellation and $(\widetilde{m}_{\nu}^{-1})_{33}=0$ (in this case the validity of Eqs.~(\ref{Mi}) breaks down).
 This would correspond to one of the two so-called crossing level solutions (the second is realised for $(\widetilde{m}_{\nu})_{11}=0$ and in that case one has $M_1=M_2$) 
 studied in \cite{Akhmedov:2003dg} for small mixing angles in $V_L$. Therefore, the fit we found is just in the vicinity of the crossing level solution $(\widetilde{m}_{\nu}^{-1})_{33}=0$
 (corresponding to $M_2=M_3$).}

Finally, let us say that an analogous result holds for inverted ordering, though this possibility is now strongly disfavoured by recent results from neutrino oscillation experiments \cite{Santos:2024eko}
and new cosmological observations placing a very stringent upper bound on the sum of neutrino masses \cite{DESI:2024mwx}.

\section{Gauge coupling unification and proton decay}\label{sec:GCU}
Proton decay is the smoking gun signal of grand unification. For a recent review on this subject, see, for example, Ref.~\cite{Dev:2022jbf}. In the framework of non-supersymmetric $SO(10)$ GUTs, the most significant proton decay mode is the $p\to e^+ \pi^0$ and the corresponding lifetime of the proton can be estimated as follows~\cite{FileviezPerez:2004hn,Nath:2006ut}:
\begin{align}
\tau_p = & \Bigg[g_\mathrm{GUT}^4  \frac{m_p}{32\pi}\left(1-\frac{m_{\pi^0}^2}{m_p^2}\right)^2 A_L^2   \times  
\nonumber \\
& 
\bigg\{ A_{SR}^2 \left(\frac{M_{X,Y}^2+M_{X^\prime,Y^\prime}^2}{M_{X,Y}^2M_{X^\prime,Y^\prime}^2} \right)^2  \left|  \langle \pi^0 \rvert (ud)_R u_L\lvert p \rangle \right|^2 
+ 
A_{SL}^2 \frac{4}{M_{X,Y}^4} \left|  \langle \pi^0 \rvert (ud)_L u_L\lvert p \rangle \right|^2 \bigg\} \Bigg]^{-1}. \label{eq:PD}
\end{align}
Here, $m_p$ and  $g_\mathrm{GUT}$ represent the proton mass and the unified gauge coupling, respectively.  $M_{X,Y}$ and $M_{X^\prime,Y^\prime}$ denote the masses of the gauge bosons $X,Y\in (3,2,5/6)$ and  $X^\prime,Y^\prime \in (3,2,-1/6)$.  For our analysis, we identify the the scale at which all gauge couplings unify, the GUT scale, with these gauge boson masses. Although the masses $M_{X,Y}$ and $M_{X^\prime,Y^\prime}$ are in general different, they are nearly degenerate in the symmetry breaking scheme that we adopt here with an intermediate Pati-Salam symmetry, as they belong to the same $(6,2,2)$ multiplet of the PS symmetry.

The values of the hadronic matrix elements appearing in Eq.~\eqref{eq:PD} have been computed on the lattice and are found to be $\langle \pi^0 \rvert (ud)_R u_L\lvert p \rangle = -0.131$ GeV$^2$ and  $\langle \pi^0 \rvert (ud)_L u_L\lvert p \rangle = 0.134$ GeV$^2$ ~\cite{Aoki:2017puj}.  Furthermore, the long-distant renormalization group running factor $A_L$ \cite{Nihei:1994tx} has a value given by $A_L\approx 1.2$, and the short range renormalization factor $A_S$ is computed as follows  \cite{Buras:1977yy,Goldman:1980ah,Caswell:1982fx,Ibanez:1984ni}: 
\begin{align} 
A_S = \prod_{j}^{M_Z\leq M_{j}\leq M_X} \prod_i \left[ \frac{\alpha_i \left(M_{j+1}\right)}{\alpha_i \left(M_{j}\right)} \right]^{\frac{\gamma_i}{b_i}} , \label{eq:anomalous}
\end{align}
where, the beta function coefficients $b_i$ are explicitly presented later in this section. Furthermore, $\alpha_i = g_i^2/4\pi $ and $\gamma_i$'s are the relevant anomalous dimensions~\cite{Abbott:1980zj}. Values of the anomalous dimensions to be used for our numerical studies are~\cite{Babu:2015bna,Chakrabortty:2019fov,Babu:2024ecl}: for the SM gauge group, $\gamma_{3C,2L,1Y}^L= \lbrace 2, \frac{9}{4}, \frac{23}{20} \rbrace$ and $\gamma_{3C,2L,1Y}^R= \lbrace 2, \frac{9}{4}, \frac{11}{20}  \rbrace$; whereas, for the Pati-Salam intermediate gauge group with parity we have $\gamma_{4C,2L,2R}^L= \gamma_{4C,2L,2R}^R= \lbrace \frac{15}{4}, \frac{9}{4}, \frac{9}{4} \rbrace$.

In order to satisfy the   the current proton lifetime bound of $\tau_p (p\to e^+\pi^0)> 2.4\times 10^{34}$ yrs from Super-Kamiokande~\cite{Super-Kamiokande:2020wjk}, one requires the GUT scale to be $M_\mathrm{GUT}\gtrsim 5\times 10^{15}$ GeV. This limit follows from Eq. (\ref{eq:PD}) when the values of the various parameters appearing in it are inserted as discussed above.  In the following we show how this limit can be respected in our scenario while also being consistent with gauge coupling unification. For this purpose a symmetry breaking scheme should be specified with the intermediate scale gauge symmetry identified.  The scalar spectrum that survive below the GUT scale down to the intermediate scale ($M_\mathrm{int}$) should also be specified. For this we shall adopt a minimal scenario dictated by the extended survival hypothesis~\cite{Georgi:1979md,delAguila:1980qag,Mohapatra:1982aq,Dimopoulos:1984ha}. Under this hypothesis, only those scalar fragments from the GUT multiplets needed for subsequent symmetry breaking are assumed to survive down to the intermediate scale.

As discussed in Sec. \ref{sec:fit}, the heaviest right-handed neutrino$N_3$  has a mass of  $2\times 10^{13}\;\mathrm{GeV} \lesssim M_{N_3} \lesssim 10^{15}\;\mathrm{GeV}$. This follows solely from the fermion fit~\cite{Saad:2022mzu}. Since $N_3$ mass can only arise after intermediate symmetry breaking, and since the relevant Yukawa coupling should be not more than order unity (so that perturbative unitarity is preserved), we infer that $(2-4)\times 10^{13}\;\mathrm{GeV} \lesssim M_\mathrm{int}$. Furthermore, the best fit for the fermion spectrum typically prefers this scale to be $M_{N_3}\sim (1-10)\times 10^{14}$ GeV, which suggests that $M_{\rm int}$ should be also in this range. Remarkably, preference of such a high intermediate scale ($M_\mathrm{int}$) points towards a particular 
symmetry breaking scheme, viz., where $SO(10)$ breaks down to the $SU(4)_c \times SU(2)_L \times SU(2)_R$ Pati-Salam subgroup with an unbroken $Z_2$ parity.  
To be specific, with the assumption of extended survival hypothesis, in Ref.~\cite{Babu:2016bmy}, two schemes of $SO(10)$ symmetry breaking was analyzed.  If a $54_H$ is used for the GUT symmetry breaking, gauge coupling unification requires $M_\mathrm{int}$ to be about two orders of magnitude below $M_\mathrm{GUT}$ (from 1-loop RGE analysis). This is fully consistent with the above-mentioned range of $2\times 10^{13}\;\mathrm{GeV} \lesssim M_\mathrm{int} \lesssim 10^{15}\;\mathrm{GeV}$.  However, if a $45_H$ or $210_H$ is used for the GUT breaking, $SO(10)$ would break down to the Pati-Salam symmetry, but without the $Z_2$ parity, in which case $M_\mathrm{int} \simeq 10^{11}$ GeV would follow~\cite{Babu:2016bmy}  (from 1-loop RGE analysis).  The fermion mass fit to the minimal Yukawa sector that we have adopted here, therefore, prefers the GUT symmetry breaking Higgs to be in the $54$-dimensional representation.

\begin{figure}[t!]
\centering
\includegraphics[width=11cm]{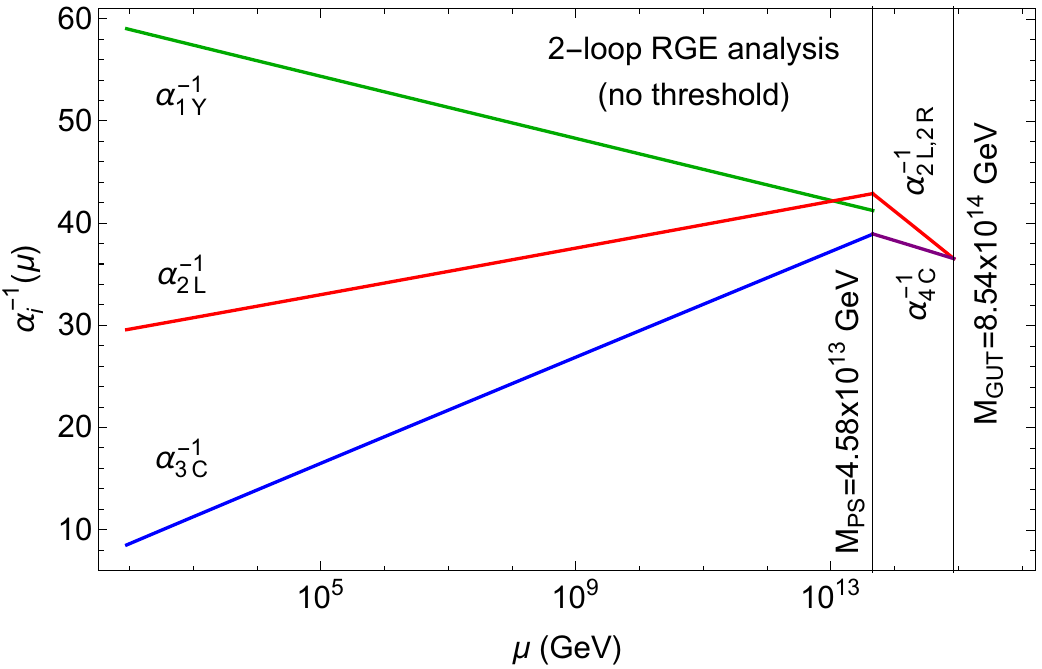}
\caption{Gauge coupling evolution with an intermediate $Su(4)_c \times SU(2)_L \times SU(2)_R$ symmetry and an unbroken $Z_2$ parity.   The minimal set of particle content affecting the running above the intermediate scale is $(1,2,2)_\mathcal{R}\subset 10_H$,  $(1,2,2)_\mathcal{R}, (15,2,2)_\mathcal{R}\subset 120_H$, and $(1,2,2), (15,2,2), (10,1,3), (\overline{10},3,1)\subset \overline{126}_H$. Here, for clarity, real representations are denoted by a subscript $\mathcal{R}$. No threshold effects are included here. All particles surviving to $M_{\rm int}$ are taken to be degenerate, and all particles at the GUT scale are also taken to be degenerate. }
\label{fig:unification}
\end{figure}

Therefore, for consistency, in this work, we consider the following symmetry breaking chain:
\begin{align}\label{eq:SSB}
SO(10) 
&\xrightarrow[54_H]{M_\mathrm{GUT}} 
SU(4)_{C}\times SU(2)_{L} \times SU(2)_{R}\times D 
\\
&\xrightarrow[126_H]{M_\mathrm{int}} 
SU(3)_{C}\times SU(2)_{L} \times U(1)_{Y} 
\\
&\xrightarrow[10_H+120_H+126_H]{M_{EW}} 
SU(3)_{C} \times U(1)_\mathrm{em} 
\end{align}
such that $M_\mathrm{GUT}> M_\mathrm{int}>M_{EW}$. As mentioned earlier, $M_\mathrm{GUT}$ is identified with the mass of the $M_{X^{(\prime)},Y^{(\prime)}}$ gauge bosons. Similarly, we  identify $M_\mathrm{int}$ with the mass of leptoquark gauge boson of the Pati-Salam symmetry,  $V_\mu (\overline 3,1,-2/3)$. The rest of the gauge bosons at $M_{\rm int}$ are not, however, degenerate with $M_V$, with their masses given by the relations $m_{W_R}/M_\mathrm{int}=g_{2R}/g_{4C}$ and $m_{Z_R}/M_\mathrm{int}=\left(3/2+g^2_{2R}/g_{4C}^2\right)^{1/2}$~\cite{Saad:2017pqj}. We shall take account of this non-degeneracy in the calculation of gauge coupling evolution.

Thus,  GUT symmetry is first spontaneously broken to the Pati-Salam symmetry with a discrete $Z_2$ symmetry $D$, which is subsequently broken down to the SM when $\overline{126}_H$ acquires a VEV of order $M_{\rm int}$. This discrete symmetry guarantees that $g_{2L,\mathrm{PS}}=g_{2R,\mathrm{PS}}$.  This symmetry breaking is also responsible for generating the mass of the heavy right-handed neutrinos.  Hereafter, with this symmetry breaking chain, we perform a detailed analysis of the gauge coupling unification. The 2-loop RGEs for the gauge couplings $g_i$ ($i=1,2,3$) are given by
\begin{align}
&\dfrac{d \alpha^{-1}_i(\mu)}{d \ln \mu}=-\dfrac{b_i}{2 \pi}-\sum\limits_{j}\dfrac{b_{ij}}{8 \pi^2 \alpha^{-1}_j(\mu)}, 
\end{align}
where $\mu$ denotes the energy scale. At the low scale, i.e., $\mu=M_Z$, we use the following values of the gauge coupling~\cite{Antusch:2013jca}: 
\begin{align}
g_1(M_Z)=   0.461425_{-0.000043}^{+0.000044},\;\;
g_2(M_Z)=   0.65184_{-0.00017}^{+0.00018},\;\;
g_3(M_Z)=   1.2143_{-0.0035}^{+0.0036}.
\end{align}
Using the generalized formula for the beta-function,  $\beta = \mu d g/d\mu$, the 1-loop and 2-loop coefficients are given by~\cite{Machacek:1983tz}
\begin{align}
b_i&=-\dfrac{11}{3}C_2 (G_i)+\dfrac{4}{3}\kappa S_2(F_i)+\dfrac{1}{6}\eta S_2(S_i),\\
b_{ij}=&-\dfrac{34}{3} \left[ C_2(G_i)\right] ^2 \delta_{ij}+\kappa\left[4 C_2(F_j) +\dfrac{20}{3}\delta_{ij} C_2(G_i)\right]S_2(F_i) \nonumber\\ 
&+\eta \left[2C_2(S_j)+\dfrac{1}{3} \delta_{ij}  C_2(G_i)\right]S_2(S_i). \end{align}
Here, \( S_2 \) and \( C_2 \) denote the Dynkin indices and quadratic Casimir of a given representation, along with their respective multiplicity factors. For Dirac (Weyl) fermion, \( \kappa = 1 \) (\( \kappa = \frac{1}{2} \)). The parameter \( \eta \) takes values \( 1 \) for real scalar fields and \( 2 \) for complex scalar fields. The symbols \( G \), \( F \), and \( S \) represent gauge multiplets, fermions, and scalars, respectively.

At the GUT scale, the appropriate matching conditions read~\cite{Weinberg:1980wa,Hall:1980kf} 
\begin{align}
&\alpha^{-1}_{4C,\mathrm{PS}}  = \alpha^{-1}_{2L,\mathrm{PS}}  =\alpha^{-1}_{2R,\mathrm{PS}}   =\alpha^{-1}_{\mathrm{GUT}},    
\end{align}
whereas, at the PS scale, we have
\begin{align}
&\alpha^{-1}_{3C,\mathrm{SM}}  = \alpha^{-1}_{4C,\mathrm{PS}}-\frac{1}{12\pi},\\ 
&\alpha^{-1}_{2L,\mathrm{SM}}  = \alpha^{-1}_{2L,\mathrm{PS}},\\ 
&\alpha^{-1}_{1Y,\mathrm{SM}}= \frac{3}{5}\left( \alpha^{-1}_{2R,\mathrm{PS}}- \frac{1}{6\pi} \right) + \frac{2}{5}\left( \alpha^{-1}_{4C,\mathrm{PS}}- \frac{1}{3\pi} \right). 
\end{align}

The well-known beta function coefficients of the SM are as follows:
\begin{align}
\left(b^\textrm{SM}_{1Y}, b^\textrm{SM}_{2L}, b^\textrm{SM}_{3C}\right)=\left(\frac{41}{10},-\frac{19}{6},-7\right), 
\;\;\;
b^\textrm{SM}_{ij}=\left(
\begin{array}{ccc}
 \frac{199}{50} & \frac{27}{10} & \frac{44}{5} \\
 \frac{9}{10} & \frac{35}{6} & 12 \\
 \frac{11}{10} & \frac{9}{2} & -26 \\
\end{array}
\right).
\end{align}
If the extended survival hypothesis is assumed (as we do), the minimal set of scalar fields that affects the running above the intermediate scale is $(1,2,2)_\mathcal{R}\subset 10_H$,  $(1,2,2)_\mathcal{R}, (15,2,2)_\mathcal{R}\subset 120_H$, and $(1,2,2), (15,2,2), (10,1,3), (\overline{10},3,1)\subset \overline{126}_H$. Here, for clarity, real representations are denoted by a subscript $\mathcal{R}$.  The $(10,1,3) \subset \overline{126}_H$ must survive down to $M_{\rm int}$ as it is responsible for the PS symmetry breaking.  
The $(\overline{10},3,1)$ fragment should then be also at $M_{\rm int}$ since it is the $Z_2$ parity partner of $(\overline{10},3,1)$, with the discrete $Z_2$ parity broken only at $M_{\rm int}$. 
The light Higgs doublet of the SM is a well-tempered mixture of the Higgs doublets from the other fragments listed above, all of which should survive down to $M_{\rm int}$.  With this particle content, we obtain the following beta function coefficients effective in the momentum range $M_{\rm int} \leq \mu \leq M_{\rm GUT}$:
\begin{align}
\left(b^\textrm{PS}_{2L}, b^\textrm{PS}_{2R}, b^\textrm{PS}_{4C}\right)=\left( \frac{23}{2},\frac{23}{2},\frac{10}{3} \right), 
\;\;\;
b^\textrm{PS}_{ij}=\left(
\begin{array}{ccc}
 \frac{593}{2} & \frac{147}{2} & \frac{1485}{2} \\
 \frac{147}{2} & \frac{593}{2} & \frac{1485}{2} \\
 \frac{297}{2} & \frac{297}{2} & \frac{4447}{6} \\
\end{array}
\right).
\end{align}
We have plotted in Fig.~\ref{fig:unification} the evolution of the gauge couplings with energy in this scenario using the full two-loop evolution equations with the assumption that all particles at the intermediate scale have a common mass, and all GUT scale particles also have a common mass. As can be seen from Fig.~\ref{fig:unification}, the GUT scale obtained with these assumptions, $M_{\rm GUT} = 8.54 \times 10^{14}$ GeV, is not large enough to evade the current proton decay limits from  Super-Kamiokande. Utilizing Eq.~\eqref{eq:PD} and using $g_\mathrm{GUT}=0.586$ corresponding to Fig.~\ref{fig:unification}, we obtain $\tau_p=8.4\times 10^{30}$ yrs, which is too rapid.

\begin{figure}[t!]
\centering
\includegraphics[width=11cm]{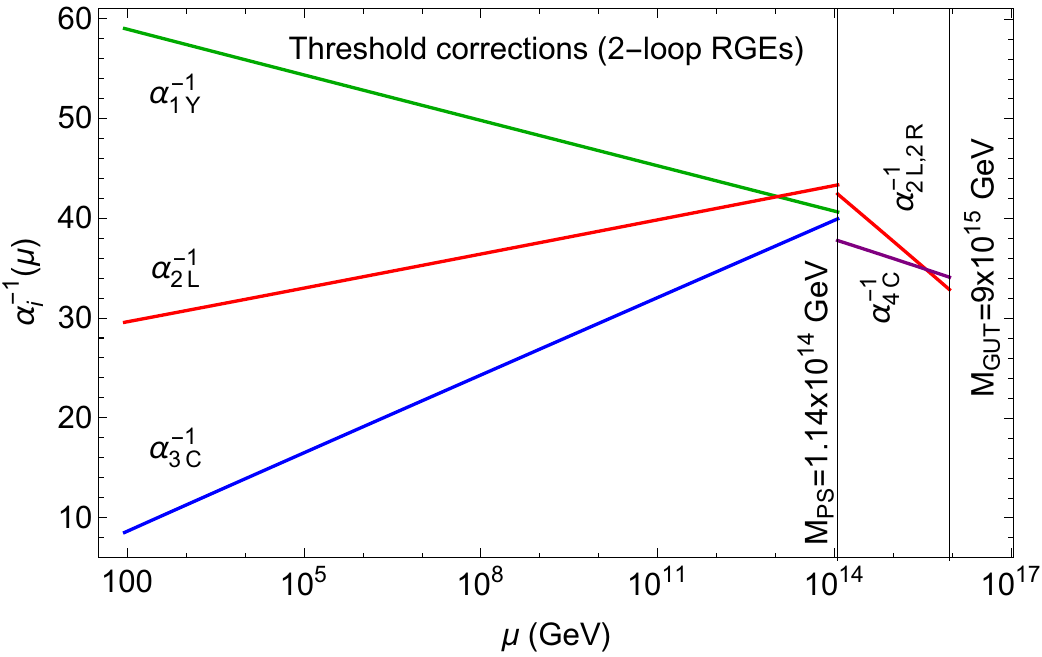}
\caption{Example of 2-loop gauge coupling unification including 1-loop threshold corrections; see text for details.  }
\label{fig:threshold}
\end{figure}

Our assumption in Fig.~\ref{fig:unification} that all GUT scale particles have a commons mass, and similarly, all intermediate scale particles are also degenerate is in fact too simplistic.  The intermediate scale gauge boson spectrum discussed earlier shows clearly that the gauge bosons don't have a common mass, although their masses are close to each other.  Similarly, a full Higgs potential analysis would reveal that the scalar sub-multiplets have differing masses, which holds for the particles at the GUT scale as well as at the intermediate scale.  A more realistic scenario would allow for small deviations from degeneracy for these masses. These threshold corrects~~\cite{Weinberg:1980wa,Hall:1980kf} at $M_{\rm GUT}$ and $M_{\rm int}$ will play a crucial role in the gauge coupling evolution, in particular, in the numerical determination of $M_{\rm GUT}$. 
Therefore, in search of a viable scenario, we consider a more general scheme, where the particle content is taken to be similar to the minimal scheme, with only the scalar states  $(1,2,2)_\mathcal{R}\subset 10_H$,  $(1,2,2)_\mathcal{R}, (15,2,2)_\mathcal{R}\subset 120_H$, and $(1,2,2), (15,2,2), (10,1,3), (\overline{10},3,1)\subset \overline{126}_H$ residing close to the intermediate scale. However, these states are now allowed to live within a narrow range $(0.1-2)\times M_\mathrm{int}$.  Similarly, states that were assumed to be perfectly degenerate with the GUT scale are now allowed to float within $(0.1-2)\times M_\mathrm{GUT}$.  (The upper range of 2 is used to ensure that perturbation theory remains valid~\cite{Babu:2015bna}.) 
Due to these threshold effects, the modified matching condition between the gauge couplings $\alpha_d^{-1}$ of the daughter gauge group ($G_d$) and the gauge couplings $\alpha_p^{-1}$ of the parent group ($G_p$) reads
\begin{align}
\alpha_d^{-1}(\mu)-\frac{C_2(G_d)}{12\pi}=\left(\alpha_p^{-1}(\mu) -\frac{C_2(G_p)}{12\pi}\right) - \frac{\Lambda_d(\mu)}{12\pi}, 
\end{align}
where the 1-loop threshold corrections involving the superheavy fields are given by~\cite{Weinberg:1980wa,Hall:1980kf,Bertolini:2009qj,Bertolini:2013vta,Babu:2015bna},
\begin{align}
\Lambda_d(\mu)=-21\; {\mathrm{Tr}} (t_{dV}^2 \ln\frac{m_V}{\mu})+2\;\eta\; {\mathrm{Tr}} (t_{dS}^2 \ln\frac{m_S}{\mu}) + 8\;\kappa\; {\mathrm{Tr}} (t_{dF}^2 \ln\frac{m_F}{\mu}).
\end{align}
Here, \( t_{dV} \), \( t_{dS} \), and \( t_{dF} \) are the generators for the representations of the superheavy vector, scalar, and fermion fields under \( G_d \), respectively, with \( m_V \), \( m_S \), and \( m_F \) representing their corresponding masses.

For a sample particle spectrum (benchmark given below\footnote{The threshold corrections presented here are some examples, they are not predictions of the model.  One could choose them differently, and in some cases the spectrum would be inconsistent with proton decay.}), gauge coupling unification is presented in Fig.~\ref{fig:threshold}. This plot illustrates that with the inclusion of threshold corrections in the range $(0.1,\,2)$, a high GUT scale of $M_\mathrm{GUT}\sim 10^{16}$ GeV can be obtained, which escapes the present proton decay bounds. The corresponding intermediate scale  $M_\mathrm{int}\sim 10^{14}$ GeV also aligns well with the fermion mass fit.  Again, utilizing Eq.~\eqref{eq:PD} and using $g_\mathrm{GUT}=0.615$ corresponding to Fig.~\ref{fig:threshold}, we obtain $\tau_p=7\times 10^{34}$ yrs, which can be probed by the 10-years of Hyper-Kamiokande~\cite{Hyper-Kamiokande:2018ofw} operations (note that the future sensitivity of Hyper-K in its 10(20)-years of operations is $7.8\times 10^{34}$ ($1.56\times 10^{35}$) yrs at $90\%$ confidence level).  The benchmark point corresponding to Fig.~\ref{fig:threshold}  has the following threshold corrections at the GUT scale:
\begin{align}
&\left( r^{(6,1,1)}_{10},  r^{(1,3,3)}_{54}, r^{(20^\prime,1,1)}_{54},   r^{(6,3,1),(6,1,3)}_{54} \right)=
(0.11, 1.93, 0.10, 1.94),
\\&
\left( r^{(10,1,1),(\overline{10},1,1)}_{120},  r^{(6,1,1)}_{\overline{126}} \right)=
(0.10,  0.11). 
\end{align}
Here we have defined $r^X_Y\equiv M^X_Y/\mu$, where $M^X_Y$ denotes the mass of the associated sub-multiplet and $\mu = M_{\rm GUT}$ for GUT-scale threshold effects and $\mu = M_{\rm int}$ for intermediate-scale threshold effects.  We assume that the masses of these sub-multiplets are independent, except for the relations dictated by the $Z_2$-parity symmetry.  This appears to be justified, since the scalar potential of the model has sufficient number of free parameters. For GUT-scale threshold effects, $X$ in 
$r^X_Y$ represents the PS sub-multiplet residing in a GUT multiplet $Y$.  The benchmark point also takes the threshold effects at $M_{\rm int}$ to be
\begin{align}
\left( r^{(3,2,7/6)}_{120}, r^{(3,2,-1/6)}_{120} , r^{(8,2,1/2)}_{120}, r^{(1,3,1)}_{\overline{126}} \right)  &=(1.73, 0.153, 0.102, 1.90),
\nonumber\\
\left( r^{(\overline 3,3,1/3)}_{\overline{126}}, r^{(\overline 6,3,-1/3)}_{\overline{126}},  r^{(1,1,-2)}_{\overline{126}}, r^{(3,1,2/3)}_{\overline{126}} \right) &=(1.95, 0.985, 1.82, 0.108),
\nonumber\\
\left( r^{(3,1,-1/3)}_{\overline{126}}, r^{(3,1,-4/3)}_{\overline{126}}, r^{(6,1,4/3)}_{\overline{126}}, r^{(6,1,1/3)}_{\overline{126}} \right) &=(0.121, 1.99, 1.40, 0.102),
\nonumber\\
\left( r^{(6,1,-2/3)}_{\overline{126}}, r^{(3,2,7/6)}_{\overline{126}}, r^{(3,2,-1/6)}_{\overline{126}}, r^{(8,2,1/2)}_{\overline{126}} \right) &=(0.103, 1.84, 0.112, 0.105).   
\end{align}
Here $\mu=M_\mathrm{int}$ and $x$ in $r^X_Y$ represents the SM sub-multiplet $X$ residing in a GUT multiplet $Y$.
The explicit form of the GUT scale threshold corrections are given as
\begin{align}
\Lambda_{2L,\mathrm{PS}}&=6\ln\left( r^{(1,3,3)}_{54} \right) +12\ln\left( r^{(6,3,1)}_{54} \right),
\\
\Lambda_{2R,\mathrm{PS}}&=6\ln\left( r^{(1,3,3)}_{54} \right) +12\ln\left( r^{(6,1,3)}_{54} \right),
\\
\Lambda_{4C,\mathrm{PS}}&= \ln\left( r^{(6,1,1)}_{10} \right)+3 \ln\left( r^{(10,1,1)}_{120} \right)+ 3\ln\left( r^{(\overline{10},1,1)}_{120} \right)+ 8\ln\left( r^{(20^\prime,1,1)}_{54} \right) 
\nonumber\\&+
3\ln\left( r^{(6,3,1)}_{54} \right)+3 \ln\left( r^{(6,1,3)}_{54} \right)+2 \ln\left( r^{(6,1,1)}_{\overline{126}} \right).
\end{align}
Similarly,  the threshold corrections at PS scale are given by
\begin{align}
\frac{5}{3} \Lambda_{1Y,\mathrm{SM}}&= \frac{49}{3} \ln\left( r^{(3,2,7/6)}_{120} \right)+\frac{1}{3} \ln\left( r^{(3,2,-1/6)}_{120} \right)+8 \ln\left( r^{(8,2,1/2)}_{120} \right)+6 \ln\left( r^{(1,3,1)}_{\overline{126}} \right) 
\nonumber\\&+
2\ln\left( r^{(\overline 3,3,1/3)}_{\overline{126}} \right)+4 \ln\left( r^{(\overline 6,3,-1/3)}_{\overline{126}} \right)+ 8\ln\left( r^{(1,1,-2)}_{\overline{126}} \right)+\frac{8}{3} \ln\left( r^{(3,1,2/3)}_{\overline{126}} \right)
\nonumber\\&+\frac{2}{3}
\ln\left( r^{(3,1,-1/3)}_{\overline{126}} \right)+ \frac{32}{3}\ln\left( r^{(3,1,-4/3)}_{\overline{126}} \right)+\frac{64}{3} \ln\left( r^{(6,1,4/3)}_{\overline{126}} \right)+\frac{4}{3} \ln\left( r^{(6,1,1/3)}_{\overline{126}} \right)
\nonumber\\&+
\frac{16}{3}\ln\left( r^{(6,1,-2/3)}_{\overline{126}} \right)+\frac{49}{3} \ln\left( r^{(3,2,7/6)}_{\overline{126}} \right)+\frac{1}{3} \ln\left( r^{(3,2,-1/6)}_{\overline{126}} \right)+8 \ln\left( r^{(8,2,1/2)}_{\overline{126}} \right)
\nonumber\\& -
21\times 6 \ln\left( \frac{g_{2R}^\mathrm{PS}}{g_{4C}^\mathrm{PS}} \right)
,
\\
\Lambda_{2L,\mathrm{SM}}&= 3\ln\left( r^{(3,2,7/6)}_{120} \right)+3 \ln\left( r^{(3,2,-1/6)}_{120} \right)+8 \ln\left( r^{(8,2,1/2)}_{120} \right)+4 \ln\left( r^{(1,3,1)}_{\overline{126}} \right) 
\nonumber\\&
+12
\ln\left( r^{(\overline 3,3,1/3)}_{\overline{126}} \right)
+24 \ln\left( r^{(\overline 6,3,-1/3)}_{\overline{126}} \right) 
+
3 \ln\left( r^{(3,2,7/6)}_{\overline{126}} \right)+ 3\ln\left( r^{(3,2,-1/6)}_{\overline{126}} \right)
\nonumber\\&
+ 8\ln\left( r^{(8,2,1/2)}_{\overline{126}} \right),
\\
\Lambda_{3C,\mathrm{SM}}&=2 \ln\left( r^{(3,2,7/6)}_{120} \right)+2 \ln\left( r^{(3,2,-1/6)}_{120} \right)+12 \ln\left( r^{(8,2,1/2)}_{120} \right)
+
3\ln\left( r^{(\overline 3,3,1/3)}_{\overline{126}} \right)
\nonumber\\&
+15 \ln\left( r^{(\overline 6,3,-1/3)}_{\overline{126}} \right)+  \ln\left( r^{(3,1,2/3)}_{\overline{126}} \right)
+
\ln\left( r^{(3,1,-1/3)}_{\overline{126}} \right)+ \ln\left( r^{(3,1,-4/3)}_{\overline{126}} \right)
\nonumber\\&
+ 5\ln\left( r^{(6,1,4/3)}_{\overline{126}} \right)+ 5\ln\left( r^{(6,1,1/3)}_{\overline{126}} \right)
+
5\ln\left( r^{(6,1,-2/3)}_{\overline{126}} \right)+ 2\ln\left( r^{(3,2,7/6)}_{\overline{126}} \right)
\nonumber\\&
+ 2\ln\left( r^{(3,2,-1/6)}_{\overline{126}} \right)+ 12\ln\left( r^{(8,2,1/2)}_{\overline{126}} \right).
\end{align}

Finally, summarizing the results obtained in this section, we have shown that with the assumption of extended survival hypothesis without allowing for threshold corrections, proton decay is too rapid, therefore does not lead to a  viable scenario. However, there is no a priori reason for all states to be perfectly degenerate in mass with the corresponding scales, namely $M_\mathrm{GUT}$ and $M_\mathrm{int}$.  Once these threshold corrections are included by assuming that the associated states can reside within a factor of $(0.1-2)$ of the corresponding scales, proton lifetime is found to be fully consistent with current limits.  Interestingly, the model prefers the lifetime to be on the lower end, and the upcoming Hyper-Kamiokande experiment  is expected to probe such a scenario.

\section{Conclusions}\label{sec:conclusions}

In this work, we have presented a simultaneous fit to the fermion masses and mixings, including neutrino oscillation parameters, and the baryon asymmetry of the Universe generated by leptogeneis, in the context of $SO(10)$ grand unified theory with a very minimal Yukawa sector.  The model employs three scalar fields taking part in the fermion mass generation: a real $10_H$, a real $120_H$ and a complex $\overline{126}_H$.  This is the simplest Higgs system in a renormalizable $SO(10)$ theory which does not rely of exterior symmetries such as $U(1)$ or $CP$ to reduce the fermion sector parameters.  The framework has 15 moduli and 12 phases that enters into the fermion masses and mixings.

We have shown that excellent fits to the fermion observables can be realized with $\chi^2 = 1.45$ for the normal ordering of neutrino masses (NO) and $\chi^2 = 5.76$ for the inverted ordering (IO) case, while also generating the baryon asymmetry with the correct sign and the right value.  Interestingly, the mass spectrum of the right-handed neutrinos in both NO and IO cases is found from the fermion fits to be $\left(M_1, M_2, M_3\right)\sim     \left(10^{4-5}, 10^{11-12}, 10^{14-15}\right)$ GeV, which leads to a scenario where $B-L$ asymmetry is dominantly produced from $N_2$ dynamics while $N_1$ is responsible for erasing the excess asymmetry. For the best-fit points, the efffective mass for neutrinoless double beta decay parameter is 3.68 (34.36) meV for NO (IO) case, consistent with the current experimental limit from KamLAND-Zen (corresponding to upper limits on $\langle m_{\beta\beta}\rangle$ of 36-156 meV~\cite{KamLAND-Zen:2022tow}) while the IO scenario can be completely probed in the future nEXO experiment~\cite{nEXO:2021ujk}. For the case of normal ordering of neutrino masses, the model  prefers the leptonic CP-violating phase to be in the range $\delta_\mathrm{CP}\simeq (230-300)^\circ$. Moreover, we have found that the fermion mass fit prefers a high scale symmetry breaking triggered by a $54_H$-dimensional representation. Our detailed numerical study reveals that non-zero threshold corrections are vital in realizing gauge coupling unification while being consistent with proton decay bounds. As long as these threshold correctins remain small, proton lifetime cannot exceed $10^{35}$ years, which may be within reach of forthcoming experiments.

\subsection*{Acknowledgments}
The work of KSB is supported in part by the U.S. Department of Energy under grant number DE-SC0016013. 
CSF acknowledges the support by Fundacão de Amparo à Pesquisa do Estado de São Paulo (FAPESP) Contracts No. 2019/11197-6 and 2022/00404-3 and Conselho Nacional de Desenvolvimento Científico e Tecnológico (CNPq) under Contract No. 304917/2023-0. KB, CSF and SS  would also like to thanks the hospitality of Fermilab in the period from May to September, 2017 (with CSF supported by FAPESP 2017/02747-7) where this work was first initiated. They also acknowledge the Center for Theoretical Underground Physics and Related Areas (CETUP* 2024) and the Institute for Underground Science at Sanford Underground Research Facility (SURF) for providing a conducive environment for the finalization of this work. 
PDB wishes to thank Xubin Hu for useful discussions and acknowledges  support from the
Munich Institute for Astro-Particle and BioPhysics (MIAPbP) funded by the Deutsche 
Forschungsgemeinschaft (DFG, German Research Foundation) under 
Germany's Excellence Strategy – EXC-2094 – 390783311.

\begin{appendices}
\section{Fit parameters: Normal ordering}\label{A}
In this Appendix we provide the fit parameters defined in Eqs.~\eqref{cR}-\eqref{E1} for the normal ordering (NO) solution. The best fit values of these parameters at the GUT scale
 are as follows: 
\begin{align}
&r_1 = -3.23303\times 10^{-3} +6.51103\times 10^{-6} i, \label{fit1} \\
&r_2= -0.355963 + 1.25289 i, \\
&\phi=-1.35108,  \\
&c_R= 9.84658\times 10^{12}, 
\end{align}

\begin{equation}
S= \left(
\begin{array}{ccc}
 6.67309\times 10^{-9} & 0 & 0 \\
 0 & 0.211339 & 0 \\
 0 & 0 & 82.3132 \\
\end{array}
\right) \;\rm{GeV},
\end{equation}

\begin{equation}
D=\scalemath{0.78}
{ 
\left(
\begin{array}{ccc}
 -2.89062\times 10^{-4}+4.87101\times 10^{-4} i & 2.34361\times
   10^{-3}+3.43972\times 10^{-3} i & -5.44162\times
   10^{-3}+3.42227\times 10^{-4} i \\
 2.34361\times 10^{-3}+3.43972\times 10^{-3} i & 4.61704\times
   10^{-2}-1.39879\times 10^{-3} i & -4.37217\times
   10^{-3}-5.40885\times 10^{-1} i \\
 -5.44162\times 10^{-3}+3.42227\times 10^{-4} i & -4.37217\times
   10^{-3}-5.40885\times 10^{-1} i & 3.83674\times
   10^{-1}+2.9757\times 10^{-2} i \\
\end{array}
\right)
}\;\rm{GeV},
\end{equation}

\begin{equation}
A=\scalemath{0.78}
{
\left(
\begin{array}{ccc}
 0 & 1.10335\times 10^{-3}+2.93066\times 10^{-3} i & -4.7391\times
   10^{-3}-1.62764\times 10^{-4} i \\
 -1.10335\times 10^{-3}-2.93066\times 10^{-3} i & 0 & 5.35481\times
   10^{-1}-1.00172\times 10^{-1} i \\
 4.7391\times 10^{-3}+1.62764\times 10^{-4} i & -5.35481\times
   10^{-1}+1.00172\times 10^{-1} i & 0 \\
\end{array}
\right)
} \;\rm{GeV}. \label{fit2}
\end{equation}

\section{Fit parameters: Inverted ordering}\label{B}
In this Appendix we provide the fit parameters defined in Eqs.~\eqref{cR}-\eqref{E1} for the inverted ordering (IO) solution. The best fit values of these parameters at the GUT scale
 are as follows: 
\begin{align}
&r_1 = 3.72884\times 10^{-3}+3.49392\times 10^{-5} i, \label{fit3} \\
&r_2= 0.146324\, +1.11548 i, \\
&\phi=-1.55817,  \\
&c_R= 7.01764\times 10^{12}, 
\end{align}

\begin{equation}
S= \left(
\begin{array}{ccc}
 1.51614\times 10^{-9} & 0 & 0 \\
 0 & 2.45908\times 10^{-1} & 0 \\
 0 & 0 & 8.34044\times 10^1 \\
\end{array}
\right) \;\rm{GeV},
\end{equation}

\begin{equation}
D=\scalemath{0.78}
{
\left(
\begin{array}{ccc}
 -7.18337\times 10^{-4}+2.32015\times 10^{-4} i & 2.67684\times
   10^{-3}+1.98135\times 10^{-3} i & -3.98144\times
   10^{-3}+8.5517\times 10^{-4} i \\
 2.67684\times 10^{-3}+1.98135\times 10^{-3} i & 6.6873\times
   10^{-3}-4.28609\times 10^{-2} i & -4.34152\times
   10^{-2}-5.38267\times 10^{-1} i \\
 -3.98144\times 10^{-3}+8.5517\times 10^{-4} i & -4.34152\times
   10^{-2}-5.38267\times 10^{-1} i & -3.33469\times
   10^{-1}-3.11386\times 10^{-2} i \\
\end{array}
\right)
}\;\rm{GeV},
\end{equation}

\begin{equation}
A=\scalemath{0.78}
{
\left(
\begin{array}{ccc}
 0 & 2.69205\times 10^{-3}+1.37192\times 10^{-3} i & -3.46624\times
   10^{-3}+7.82184\times 10^{-4} i \\
 -2.69205\times 10^{-3}-1.37192\times 10^{-3} i & 0 & 5.54677\times
   10^{-1}-3.55935\times 10^{-2} i \\
 3.46624\times 10^{-3}-7.82184\times 10^{-4} i & -5.54677\times
   10^{-1}+3.55935\times 10^{-2} i & 0 \\
\end{array}
\right)
} \;\rm{GeV}. \label{fit4}
\end{equation}

\section{Parameters for leptogenesis}\label{C}
For the convenience of the reader, here we present the neutrino Dirac, Eq.~\eqref{MDnu}, and the charged lepton, Eq.~\eqref{ME}, Yukawa coupling matrices in a basis where the right-handed neutrino mass matrix, Eq.~\eqref{E2}, is diagonal. These quantities are given at the $M_2$ scale used for leptogenesis and at $M_1$ scale used for washout calculations. Notice that in our flavor-covariant formalism, it is not necessary to use the charged lepton mass diagonal basis. As we have shown in Sec. \ref{sec:lepto}, $B$ or $B-L$ asymmetry is invariant under flavor rotations, see Eq. \eqref{eq:flavor_rotations}.
These Yukawa coupling and masses quoted below are the outcomes of the fits presented in Appendices.~\ref{A} and~\ref{B}. At renormalization scale $M_2$, we have for
\begin{align}
&\mathrm{normal\; ordering\; (NO):}
\nonumber\\
&
\left(M_1, M_2, M_3\right)= \left( 6.57071\times 10^4, 2.08095\times 10^{12}, 7.51307\times 10^{14} \right)\; \mathrm{GeV}, \label{eq:117}
\\
&y_{\nu_D}=\scalemath{0.78}
{
\left(
\begin{array}{ccc}
 -1.60399\times 10^{-6}+2.70269\times 10^{-6} i & 3.83439\times
   10^{-6}-1.6047\times 10^{-6} i & 3.99098\times
   10^{-6}+1.09775\times 10^{-6} i \\
 2.21767\times 10^{-5}+3.97769\times 10^{-5} i & -3.26201\times
   10^{-3}-7.64731\times 10^{-6} i & -3.77518\times
   10^{-3}-2.01867\times 10^{-3} i \\
 -6.07975\times 10^{-5}+2.51138\times 10^{-6} i & 3.58336\times
   10^{-3}-3.71946\times 10^{-3} i & -1.29043+1.55752\times
   10^{-4} i \\
\end{array}
\right),
}
\\
&y_E=\scalemath{0.78}
{
\left(
\begin{array}{ccc}
 -1.6569\times 10^{-6}+2.79187\times 10^{-6} i & -9.85534\times
   10^{-6}+2.16613\times 10^{-5} i & -2.15379\times
   10^{-5}-3.35986\times 10^{-5} i \\
 3.67292\times 10^{-5}+1.77743\times 10^{-5} i & 2.76287\times
   10^{-4}-8.26329\times 10^{-6} i & -4.2159\times
   10^{-4}+1.00513\times 10^{-3} i \\
 -4.20094\times 10^{-5}+3.56618\times 10^{-5} i & 3.46994\times
   10^{-4}-7.14891\times 10^{-3} i & 7.17039\times
   10^{-3}+1.71882\times 10^{-4} i \\
\end{array}
\right).
}
\\
&\mathrm{and for inverted\; ordering\; (IO):}
\nonumber\\
&
\left(M_1, M_2, M_3\right)= \left( 1.06397\times 10^4, 1.72569\times 10^{12}, 5.37463\times 10^{14} \right)\; \mathrm{GeV}, 
\\
&y_{\nu_D}=\scalemath{0.78}
{
\left(
\begin{array}{ccc}
 -3.96655\times 10^{-6}+1.28114\times 10^{-6} i & -5.56661\times
   10^{-7}+1.20903\times 10^{-7} i & 2.06712\times
   10^{-8}+2.91964\times 10^{-6} i \\
 3.01215\times 10^{-5}+2.17624\times 10^{-5} i & -4.03696\times
   10^{-3}-2.36699\times 10^{-4} i & -3.60534\times
   10^{-3}-3.17552\times 10^{-3} i \\
 -4.12295\times 10^{-5}+6.06392\times 10^{-6} i & 3.00083\times
   10^{-3}-2.53226\times 10^{-3} i & -1.29604-1.61077\times
   10^{-4} i \\
\end{array}
\right),
}
\\
&y_E=\scalemath{0.78}
{
\left(
\begin{array}{ccc}
 -4.10011\times 10^{-6}+1.32424\times 10^{-6} i & 8.79887\times
   10^{-6}+2.95932\times 10^{-5} i & -3.26182\times
   10^{-5}-1.76321\times 10^{-5} i \\
 2.1766\times 10^{-5}-6.97592\times 10^{-6} i & 2.23477\times
   10^{-5}-2.44794\times 10^{-4} i & 4.71227\times
   10^{-4}+4.57941\times 10^{-4} i \\
 -1.4865\times 10^{-5}+2.62983\times 10^{-5} i & -9.36539\times
   10^{-4}-6.5749\times 10^{-3} i & -7.70666\times
   10^{-3}-2.41542\times 10^{-4} i \\
\end{array}
\right).
}
\end{align}

At the scale $M_1$, we have $M_i$ as in Eq. (\ref{eq:117}) and 
\begin{align}
&\mathrm{NO:}
\nonumber\\
&y_{\nu_D}=\scalemath{0.78}
{
\left(
\begin{array}{ccc}
 -1.55316\times 10^{-6}+2.61706\times 10^{-6} i & 3.71295\times 10^{-6}-1.55390\times 10^{-6} i &
   3.86458\times 10^{-6}+1.06301\times 10^{-6} i \\
 0.0000221767\, +0.0000397769 i & -0.00326201-7.64730\times 10^{-6} i & -0.00377518-0.00201867 i \\
 -0.0000607975+2.51138\times 10^{-6} i & 0.00358336\,
   -0.00371946 i & -1.29043+0.000155752 i \\
\end{array}
\right),
}
\\
&y_E=\scalemath{0.78}
{
\left(
\begin{array}{ccc}
 -1.68764\times 10^{-6}+2.84365\times 10^{-6} i & -0.0000100378+0.0000220627 i & -0.000021937-0.0000342215 i \\
 0.0000374106\, +0.0000181041 i & 0.0002814\, -8.42045\times 10^{-6} i & -0.000429407+0.00102376 i \\
 -0.0000427879+0.0000363228 i & 0.000353425\, -0.00728139 i & 0.00730327\, +0.000175067
   i \\
\end{array}
\right).
}
\\
&\mathrm{IO:}
\nonumber\\
&y_{\nu_D}=\scalemath{0.78}
{
\left(
\begin{array}{ccc}
 -3.80357\times 10^{-6}+1.2285\times 10^{-6} i & -5.33817\times
   10^{-7}+1.15942\times 10^{-7} i & 1.98218\times
   10^{-8}+2.79971\times 10^{-6} i \\
 3.01215\times 10^{-5}+2.17624\times 10^{-5} i & -4.03696\times
   10^{-3}-2.36699\times 10^{-4} i & -3.60534\times
   10^{-3}-3.17552\times 10^{-3} i \\
 -4.12295\times 10^{-5}+6.06392\times 10^{-6} i & 3.00083\times
   10^{-3}-2.53226\times 10^{-3} i & -1.29604-1.61077\times
   10^{-4} i \\
\end{array}
\right),
}
\\
&y_E=\scalemath{0.78}
{
\left(
\begin{array}{ccc}
 -4.15257\times 10^{-6}+1.34119\times 10^{-6} i & 8.91127\times
   10^{-6}+2.99719\times 10^{-5} i & -3.30355\times
   10^{-5}-1.78575\times 10^{-5} i \\
 2.20445\times 10^{-5}-7.06517\times 10^{-6} i & 2.26371\times
   10^{-5}-2.4793\times 10^{-4} i & 4.77254\times
   10^{-4}+4.63796\times 10^{-4} i \\
 -1.50549\times 10^{-5}+2.66343\times 10^{-5} i & -9.48504\times
   10^{-4}-6.6589\times 10^{-3} i & -7.80512\times
   10^{-3}-2.44628\times 10^{-4} i \\
\end{array}
\right).
}
\end{align}

\end{appendices}
\bibliographystyle{style}
\bibliography{reference}

\providecommand{\href}[2]{#2}\begingroup\raggedright\begin{thebibliography}{100}

\bibitem{Pati:1974yy}
J.~C. Pati and A.~Salam, \emph{{Lepton Number as the Fourth Color}},
  \href{https://doi.org/10.1103/PhysRevD.10.275}{\emph{Phys. Rev. D} {\bfseries
  10} (1974) 275}.

\bibitem{Georgi:1974sy}
H.~Georgi and S.~Glashow, \emph{{Unity of All Elementary Particle Forces}},
  \href{https://doi.org/10.1103/PhysRevLett.32.438}{\emph{Phys. Rev. Lett.}
  {\bfseries 32} (1974) 438}.

\bibitem{Georgi:1974yf}
H.~Georgi, H.~R. Quinn and S.~Weinberg, \emph{{Hierarchy of Interactions in
  Unified Gauge Theories}},
  \href{https://doi.org/10.1103/PhysRevLett.33.451}{\emph{Phys. Rev. Lett.}
  {\bfseries 33} (1974) 451}.

\bibitem{Georgi:1974my}
H.~Georgi, \emph{{The State of the Art---Gauge Theories}},
  \href{https://doi.org/10.1063/1.2947450}{\emph{AIP Conf. Proc.} {\bfseries
  23} (1975) 575}.

\bibitem{Fritzsch:1974nn}
H.~Fritzsch and P.~Minkowski, \emph{{Unified Interactions of Leptons and
  Hadrons}}, \href{https://doi.org/10.1016/0003-4916(75)90211-0}{\emph{Annals
  Phys.} {\bfseries 93} (1975) 193}.

\bibitem{Pati:1973rp}
J.~C. Pati and A.~Salam, \emph{{Is Baryon Number Conserved?}},
  \href{https://doi.org/10.1103/PhysRevLett.31.661}{\emph{Phys. Rev. Lett.}
  {\bfseries 31} (1973) 661}.

\bibitem{Minkowski:1977sc}
P.~Minkowski, \emph{{$\mu \to e\gamma$ at a Rate of One Out of $10^{9}$ Muon
  Decays?}}, \href{https://doi.org/10.1016/0370-2693(77)90435-X}{\emph{Phys.
  Lett. B} {\bfseries 67} (1977) 421}.

\bibitem{Yanagida:1979as}
T.~Yanagida, \emph{{Horizontal gauge symmetry and masses of neutrinos}},
  {\emph{Conf. Proc. C} {\bfseries 7902131} (1979) 95}.

\bibitem{Glashow:1979nm}
S.~Glashow, \emph{{The Future of Elementary Particle Physics}},
  \href{https://doi.org/10.1007/978-1-4684-7197-7\_15}{\emph{NATO Sci. Ser. B}
  {\bfseries 61} (1980) 687}.

\bibitem{Gell-Mann:1979vob}
M.~Gell-Mann, P.~Ramond and R.~Slansky, \emph{{Complex Spinors and Unified
  Theories}}, {\emph{Conf. Proc. C} {\bfseries 790927} (1979) 315}
  [\href{https://arxiv.org/abs/1306.4669}{{\ttfamily arXiv:1306.4669}}].

\bibitem{Mohapatra:1979ia}
R.~N. Mohapatra and G.~Senjanovic, \emph{{Neutrino Mass and Spontaneous Parity
  Nonconservation}},
  \href{https://doi.org/10.1103/PhysRevLett.44.912}{\emph{Phys. Rev. Lett.}
  {\bfseries 44} (1980) 912}.

\bibitem{Schechter:1980gr}
J.~Schechter and J.~W.~F. Valle, \emph{{Neutrino Masses in SU(2) x U(1)
  Theories}}, \href{https://doi.org/10.1103/PhysRevD.22.2227}{\emph{Phys. Rev.}
  {\bfseries D22} (1980) 2227}.

\bibitem{Schechter:1981cv}
J.~Schechter and J.~W.~F. Valle, \emph{{Neutrino Decay and Spontaneous
  Violation of Lepton Number}},
  \href{https://doi.org/10.1103/PhysRevD.25.774}{\emph{Phys. Rev.} {\bfseries
  D25} (1982) 774}.

\bibitem{Fukugita:1986hr}
M.~Fukugita and T.~Yanagida, \emph{{Baryogenesis Without Grand Unification}},
  \href{https://doi.org/10.1016/0370-2693(86)91126-3}{\emph{Phys. Lett. B}
  {\bfseries 174} (1986) 45}.

\bibitem{Klinkhamer:1984di}
F.~R. Klinkhamer and N.~S. Manton, \emph{{A Saddle Point Solution in the
  Weinberg-Salam Theory}},
  \href{https://doi.org/10.1103/PhysRevD.30.2212}{\emph{Phys. Rev. D}
  {\bfseries 30} (1984) 2212}.

\bibitem{Arnold:1987mh}
P.~B. Arnold and L.~D. McLerran, \emph{{Sphalerons, Small Fluctuations and
  Baryon Number Violation in Electroweak Theory}},
  \href{https://doi.org/10.1103/PhysRevD.36.581}{\emph{Phys. Rev. D} {\bfseries
  36} (1987) 581}.

\bibitem{Arnold:1987zg}
P.~B. Arnold and L.~D. McLerran, \emph{{The Sphaleron Strikes Back}},
  \href{https://doi.org/10.1103/PhysRevD.37.1020}{\emph{Phys. Rev. D}
  {\bfseries 37} (1988) 1020}.

\bibitem{Buchmuller:2004nz}
W.~Buchmuller, P.~Di~Bari and M.~Plumacher, \emph{{Leptogenesis for
  pedestrians}}, \href{https://doi.org/10.1016/j.aop.2004.02.003}{\emph{Annals
  Phys.} {\bfseries 315} (2005) 305}
  [\href{https://arxiv.org/abs/hep-ph/0401240}{{\ttfamily
  arXiv:hep-ph/0401240}}].

\bibitem{Nir:2007zq}
Y.~Nir, \emph{{Introduction to leptogenesis}},  in \emph{{6th Rencontres du
  Vietnam}: {Challenges in Particle Astrophysics}}, 2, 2007,
  \href{https://arxiv.org/abs/hep-ph/0702199}{{\ttfamily
  arXiv:hep-ph/0702199}}.

\bibitem{Davidson:2008bu}
S.~Davidson, E.~Nardi and Y.~Nir, \emph{{Leptogenesis}},
  \href{https://doi.org/10.1016/j.physrep.2008.06.002}{\emph{Phys. Rept.}
  {\bfseries 466} (2008) 105}
  [\href{https://arxiv.org/abs/0802.2962}{{\ttfamily arXiv:0802.2962}}].

\bibitem{Pilaftsis:2009pk}
A.~Pilaftsis, \emph{{The Little Review on Leptogenesis}},
  \href{https://doi.org/10.1088/1742-6596/171/1/012017}{\emph{J. Phys. Conf.
  Ser.} {\bfseries 171} (2009) 012017}
  [\href{https://arxiv.org/abs/0904.1182}{{\ttfamily arXiv:0904.1182}}].

\bibitem{DiBari:2012fz}
P.~Di~Bari, \emph{{An introduction to leptogenesis and neutrino properties}},
  \href{https://doi.org/10.1080/00107514.2012.701096}{\emph{Contemp. Phys.}
  {\bfseries 53} (2012) 315} [\href{https://arxiv.org/abs/1206.3168}{{\ttfamily
  arXiv:1206.3168}}].

\bibitem{Fong:2012buy}
C.~S. Fong, E.~Nardi and A.~Riotto, \emph{{Leptogenesis in the Universe}},
  \href{https://doi.org/10.1155/2012/158303}{\emph{Adv. High Energy Phys.}
  {\bfseries 2012} (2012) 158303}
  [\href{https://arxiv.org/abs/1301.3062}{{\ttfamily arXiv:1301.3062}}].

\bibitem{Chun:2017spz}
E.~Chun et~al., \emph{{Probing Leptogenesis}},
  \href{https://doi.org/10.1142/S0217751X18420058}{\emph{Int. J. Mod. Phys. A}
  {\bfseries 33} (2018) 1842005}
  [\href{https://arxiv.org/abs/1711.02865}{{\ttfamily arXiv:1711.02865}}].

\bibitem{Dev:2017trv}
P.~S.~B. Dev, P.~Di~Bari, B.~Garbrecht, S.~Lavignac, P.~Millington and
  D.~Teresi, \emph{{Flavor effects in leptogenesis}},
  \href{https://doi.org/10.1142/S0217751X18420010}{\emph{Int. J. Mod. Phys. A}
  {\bfseries 33} (2018) 1842001}
  [\href{https://arxiv.org/abs/1711.02861}{{\ttfamily arXiv:1711.02861}}].

\bibitem{Bodeker:2020ghk}
D.~Bodeker and W.~Buchmuller, \emph{{Baryogenesis from the weak scale to the
  grand unification scale}},
  \href{https://doi.org/10.1103/RevModPhys.93.035004}{\emph{Rev. Mod. Phys.}
  {\bfseries 93} (2021) 035004}
  [\href{https://arxiv.org/abs/2009.07294}{{\ttfamily arXiv:2009.07294}}].

\bibitem{Babu:1992ia}
K.~S. Babu and R.~Mohapatra, \emph{{Predictive neutrino spectrum in minimal
  SO(10) grand unification}},
  \href{https://doi.org/10.1103/PhysRevLett.70.2845}{\emph{Phys. Rev. Lett.}
  {\bfseries 70} (1993) 2845}
  [\href{https://arxiv.org/abs/hep-ph/9209215}{{\ttfamily
  arXiv:hep-ph/9209215}}].

\bibitem{Bajc:2001fe}
B.~Bajc, G.~Senjanovic and F.~Vissani, \emph{{How neutrino and charged fermion
  masses are connected within minimal supersymmetric SO(10)}},
  \href{https://doi.org/10.22323/1.007.0198}{\emph{PoS} {\bfseries HEP2001}
  (2001) 198} [\href{https://arxiv.org/abs/hep-ph/0110310}{{\ttfamily
  arXiv:hep-ph/0110310}}].

\bibitem{Bajc:2002iw}
B.~Bajc, G.~Senjanovic and F.~Vissani, \emph{{b - tau unification and large
  atmospheric mixing: A Case for noncanonical seesaw}},
  \href{https://doi.org/10.1103/PhysRevLett.90.051802}{\emph{Phys. Rev. Lett.}
  {\bfseries 90} (2003) 051802}
  [\href{https://arxiv.org/abs/hep-ph/0210207}{{\ttfamily
  arXiv:hep-ph/0210207}}].

\bibitem{Fukuyama:2002ch}
T.~Fukuyama and N.~Okada, \emph{{Neutrino oscillation data versus minimal
  supersymmetric SO(10) model}},
  \href{https://doi.org/10.1088/1126-6708/2002/11/011}{\emph{JHEP} {\bfseries
  11} (2002) 011} [\href{https://arxiv.org/abs/hep-ph/0205066}{{\ttfamily
  arXiv:hep-ph/0205066}}].

\bibitem{Goh:2003sy}
H.~S. Goh, R.~N. Mohapatra and S.-P. Ng, \emph{{Minimal SUSY SO(10), b tau
  unification and large neutrino mixings}},
  \href{https://doi.org/10.1016/j.physletb.2003.08.011}{\emph{Phys. Lett.}
  {\bfseries B570} (2003) 215}
  [\href{https://arxiv.org/abs/hep-ph/0303055}{{\ttfamily
  arXiv:hep-ph/0303055}}].

\bibitem{Goh:2003hf}
H.~S. Goh, R.~N. Mohapatra and S.-P. Ng, \emph{{Minimal SUSY SO(10) model and
  predictions for neutrino mixings and leptonic CP violation}},
  \href{https://doi.org/10.1103/PhysRevD.68.115008}{\emph{Phys. Rev.}
  {\bfseries D68} (2003) 115008}
  [\href{https://arxiv.org/abs/hep-ph/0308197}{{\ttfamily
  arXiv:hep-ph/0308197}}].

\bibitem{Bertolini:2004eq}
S.~Bertolini, M.~Frigerio and M.~Malinsky, \emph{{Fermion masses in SUSY SO(10)
  with type II seesaw: A Non-minimal predictive scenario}},
  \href{https://doi.org/10.1103/PhysRevD.70.095002}{\emph{Phys. Rev.}
  {\bfseries D70} (2004) 095002}
  [\href{https://arxiv.org/abs/hep-ph/0406117}{{\ttfamily
  arXiv:hep-ph/0406117}}].

\bibitem{Bertolini:2005qb}
S.~Bertolini and M.~Malinsky, \emph{{On CP violation in minimal renormalizable
  SUSY SO(10) and beyond}},
  \href{https://doi.org/10.1103/PhysRevD.72.055021}{\emph{Phys. Rev. D}
  {\bfseries 72} (2005) 055021}
  [\href{https://arxiv.org/abs/hep-ph/0504241}{{\ttfamily
  arXiv:hep-ph/0504241}}].

\bibitem{Babu:2005ia}
K.~S. Babu and C.~Macesanu, \emph{{Neutrino masses and mixings in a minimal
  SO(10) model}}, \href{https://doi.org/10.1103/PhysRevD.72.115003}{\emph{Phys.
  Rev.} {\bfseries D72} (2005) 115003}
  [\href{https://arxiv.org/abs/hep-ph/0505200}{{\ttfamily
  arXiv:hep-ph/0505200}}].

\bibitem{Bertolini:2006pe}
S.~Bertolini, T.~Schwetz and M.~Malinsky, \emph{{Fermion masses and mixings in
  SO(10) models and the neutrino challenge to SUSY GUTs}},
  \href{https://doi.org/10.1103/PhysRevD.73.115012}{\emph{Phys. Rev.}
  {\bfseries D73} (2006) 115012}
  [\href{https://arxiv.org/abs/hep-ph/0605006}{{\ttfamily
  arXiv:hep-ph/0605006}}].

\bibitem{Bajc:2008dc}
B.~Bajc, I.~Dorsner and M.~Nemevsek, \emph{{Minimal SO(10) splits
  supersymmetry}},
  \href{https://doi.org/10.1088/1126-6708/2008/11/007}{\emph{JHEP} {\bfseries
  11} (2008) 007} [\href{https://arxiv.org/abs/0809.1069}{{\ttfamily
  arXiv:0809.1069}}].

\bibitem{Joshipura:2011nn}
A.~S. Joshipura and K.~M. Patel, \emph{{Fermion Masses in SO(10) Models}},
  \href{https://doi.org/10.1103/PhysRevD.83.095002}{\emph{Phys. Rev.}
  {\bfseries D83} (2011) 095002}
  [\href{https://arxiv.org/abs/1102.5148}{{\ttfamily arXiv:1102.5148}}].

\bibitem{Altarelli:2013aqa}
G.~Altarelli and D.~Meloni, \emph{{A non supersymmetric SO(10) grand unified
  model for all the physics below $M_{GUT}$}},
  \href{https://doi.org/10.1007/JHEP08(2013)021}{\emph{JHEP} {\bfseries 08}
  (2013) 021} [\href{https://arxiv.org/abs/1305.1001}{{\ttfamily
  arXiv:1305.1001}}].

\bibitem{Dueck:2013gca}
A.~Dueck and W.~Rodejohann, \emph{{Fits to SO(10) Grand Unified Models}},
  \href{https://doi.org/10.1007/JHEP09(2013)024}{\emph{JHEP} {\bfseries 09}
  (2013) 024} [\href{https://arxiv.org/abs/1306.4468}{{\ttfamily
  arXiv:1306.4468}}].

\bibitem{Fukuyama:2015kra}
T.~Fukuyama, K.~Ichikawa and Y.~Mimura, \emph{{Revisiting fermion mass and
  mixing fits in the minimal SUSY $SO(10)$ GUT}},
  \href{https://doi.org/10.1103/PhysRevD.94.075018}{\emph{Phys. Rev.}
  {\bfseries D94} (2016) 075018}
  [\href{https://arxiv.org/abs/1508.07078}{{\ttfamily arXiv:1508.07078}}].

\bibitem{Babu:2016cri}
K.~S. Babu, B.~Bajc and S.~Saad, \emph{{New Class of SO(10) Models for
  Flavor}}, \href{https://doi.org/10.1103/PhysRevD.94.015030}{\emph{Phys. Rev.
  D} {\bfseries 94} (2016) 015030}
  [\href{https://arxiv.org/abs/1605.05116}{{\ttfamily arXiv:1605.05116}}].

\bibitem{Babu:2016bmy}
K.~S. Babu, B.~Bajc and S.~Saad, \emph{{Yukawa Sector of Minimal SO(10)
  Unification}}, \href{https://doi.org/10.1007/JHEP02(2017)136}{\emph{JHEP}
  {\bfseries 02} (2017) 136}
  [\href{https://arxiv.org/abs/1612.04329}{{\ttfamily arXiv:1612.04329}}].

\bibitem{Babu:2018tfi}
K.~S. Babu, B.~Bajc and S.~Saad, \emph{{Resurrecting Minimal Yukawa Sector of
  SUSY SO(10)}}, \href{https://doi.org/10.1007/JHEP10(2018)135}{\emph{JHEP}
  {\bfseries 10} (2018) 135}
  [\href{https://arxiv.org/abs/1805.10631}{{\ttfamily arXiv:1805.10631}}].

\bibitem{Babu:2018qca}
K.~S. Babu, T.~Fukuyama, S.~Khan and S.~Saad, \emph{{Peccei-Quinn Symmetry and
  Nucleon Decay in Renormalizable SUSY $SO$(10)}},
  \href{https://doi.org/10.1007/JHEP06(2019)045}{\emph{JHEP} {\bfseries 06}
  (2019) 045} [\href{https://arxiv.org/abs/1812.11695}{{\ttfamily
  arXiv:1812.11695}}].

\bibitem{Ohlsson:2018qpt}
T.~Ohlsson and M.~Pernow, \emph{{Running of Fermion Observables in
  Non-Supersymmetric SO(10) Models}},
  \href{https://doi.org/10.1007/JHEP11(2018)028}{\emph{JHEP} {\bfseries 11}
  (2018) 028} [\href{https://arxiv.org/abs/1804.04560}{{\ttfamily
  arXiv:1804.04560}}].

\bibitem{Ohlsson:2019sja}
T.~Ohlsson and M.~Pernow, \emph{{Fits to Non-Supersymmetric SO(10) Models with
  Type I and II Seesaw Mechanisms Using Renormalization Group Evolution}},
  \href{https://doi.org/10.1007/JHEP06(2019)085}{\emph{JHEP} {\bfseries 06}
  (2019) 085} [\href{https://arxiv.org/abs/1903.08241}{{\ttfamily
  arXiv:1903.08241}}].

\bibitem{Babu:2020tnf}
K.~S. Babu and S.~Saad, \emph{{Flavor Hierarchies from Clockwork in SO(10)
  GUT}}, \href{https://doi.org/10.1103/PhysRevD.103.015009}{\emph{Phys. Rev. D}
  {\bfseries 103} (2021) 015009}
  [\href{https://arxiv.org/abs/2007.16085}{{\ttfamily arXiv:2007.16085}}].

\bibitem{Mummidi:2021anm}
V.~S. Mummidi and K.~M. Patel, \emph{{Leptogenesis and fermion mass fit in a
  renormalizable SO(10) model}},
  \href{https://doi.org/10.1007/JHEP12(2021)042}{\emph{JHEP} {\bfseries 12}
  (2021) 042} [\href{https://arxiv.org/abs/2109.04050}{{\ttfamily
  arXiv:2109.04050}}].

\bibitem{Saad:2022mzu}
S.~Saad, \emph{{Probing minimal grand unification through gravitational waves,
  proton decay, and fermion masses}},
  \href{https://doi.org/10.1007/JHEP04(2023)058}{\emph{JHEP} {\bfseries 04}
  (2023) 058} [\href{https://arxiv.org/abs/2212.05291}{{\ttfamily
  arXiv:2212.05291}}].

\bibitem{Haba:2023dvo}
N.~Haba, Y.~Shimizu and T.~Yamada, \emph{{Neutrino Mass in Non-Supersymmetric
  $SO(10)$ GUT}},
  \href{https://doi.org/10.1103/PhysRevD.108.095005}{\emph{Phys. Rev. D}
  {\bfseries 108} (2023) 095005}
  [\href{https://arxiv.org/abs/2304.06263}{{\ttfamily arXiv:2304.06263}}].

\bibitem{Kaladharan:2023zbr}
A.~Kaladharan and S.~Saad, \emph{{Fermion mass, axion dark matter, and
  leptogenesis in SO(10) GUT}},
  \href{https://doi.org/10.1103/PhysRevD.109.055010}{\emph{Phys. Rev. D}
  {\bfseries 109} (2024) 055010}
  [\href{https://arxiv.org/abs/2308.04497}{{\ttfamily arXiv:2308.04497}}].

\bibitem{Buchmuller:1996pa}
W.~Buchmuller and M.~Plumacher, \emph{{Baryon asymmetry and neutrino mixing}},
  \href{https://doi.org/10.1016/S0370-2693(96)01232-4}{\emph{Phys. Lett. B}
  {\bfseries 389} (1996) 73}
  [\href{https://arxiv.org/abs/hep-ph/9608308}{{\ttfamily
  arXiv:hep-ph/9608308}}].

\bibitem{Nezri:2000pb}
E.~Nezri and J.~Orloff, \emph{{Neutrino oscillations versus leptogenesis in
  SO(10) models}},
  \href{https://doi.org/10.1088/1126-6708/2003/04/020}{\emph{JHEP} {\bfseries
  04} (2003) 020} [\href{https://arxiv.org/abs/hep-ph/0004227}{{\ttfamily
  arXiv:hep-ph/0004227}}].

\bibitem{Buccella:2001tq}
F.~Buccella, D.~Falcone and F.~Tramontano, \emph{{Baryogenesis via leptogenesis
  in SO(10) models}},
  \href{https://doi.org/10.1016/S0370-2693(01)01409-5}{\emph{Phys. Lett. B}
  {\bfseries 524} (2002) 241}
  [\href{https://arxiv.org/abs/hep-ph/0108172}{{\ttfamily
  arXiv:hep-ph/0108172}}].

\bibitem{Branco:2002kt}
G.~C. Branco, R.~Gonzalez~Felipe, F.~R. Joaquim and M.~N. Rebelo,
  \emph{{Leptogenesis, CP violation and neutrino data: What can we learn?}},
  \href{https://doi.org/10.1016/S0550-3213(02)00478-9}{\emph{Nucl. Phys. B}
  {\bfseries 640} (2002) 202}
  [\href{https://arxiv.org/abs/hep-ph/0202030}{{\ttfamily
  arXiv:hep-ph/0202030}}].

\bibitem{Akhmedov:2003dg}
E.~K. Akhmedov, M.~Frigerio and A.~Y. Smirnov, \emph{{Probing the seesaw
  mechanism with neutrino data and leptogenesis}},
  \href{https://doi.org/10.1088/1126-6708/2003/09/021}{\emph{JHEP} {\bfseries
  09} (2003) 021} [\href{https://arxiv.org/abs/hep-ph/0305322}{{\ttfamily
  arXiv:hep-ph/0305322}}].

\bibitem{DiBari:2008mp}
P.~Di~Bari and A.~Riotto, \emph{{Successful type I Leptogenesis with
  SO(10)-inspired mass relations}},
  \href{https://doi.org/10.1016/j.physletb.2008.12.054}{\emph{Phys. Lett. B}
  {\bfseries 671} (2009) 462}
  [\href{https://arxiv.org/abs/0809.2285}{{\ttfamily arXiv:0809.2285}}].

\bibitem{DiBari:2010ux}
P.~Di~Bari and A.~Riotto, \emph{{Testing SO(10)-inspired leptogenesis with low
  energy neutrino experiments}},
  \href{https://doi.org/10.1088/1475-7516/2011/04/037}{\emph{JCAP} {\bfseries
  04} (2011) 037} [\href{https://arxiv.org/abs/1012.2343}{{\ttfamily
  arXiv:1012.2343}}].

\bibitem{Buccella:2012kc}
F.~Buccella, D.~Falcone, C.~S. Fong, E.~Nardi and G.~Ricciardi,
  \emph{{Squeezing out predictions with leptogenesis from SO(10)}},
  \href{https://doi.org/10.1103/PhysRevD.86.035012}{\emph{Phys. Rev. D}
  {\bfseries 86} (2012) 035012}
  [\href{https://arxiv.org/abs/1203.0829}{{\ttfamily arXiv:1203.0829}}].

\bibitem{DiBari:2013qja}
P.~Di~Bari and L.~Marzola, \emph{{SO(10)-inspired solution to the problem of
  the initial conditions in leptogenesis}},
  \href{https://doi.org/10.1016/j.nuclphysb.2013.10.027}{\emph{Nucl. Phys. B}
  {\bfseries 877} (2013) 719}
  [\href{https://arxiv.org/abs/1308.1107}{{\ttfamily arXiv:1308.1107}}].

\bibitem{DiBari:2014eya}
P.~Di~Bari, L.~Marzola and M.~Re~Fiorentin, \emph{{Decrypting $SO(10)$-inspired
  leptogenesis}},
  \href{https://doi.org/10.1016/j.nuclphysb.2015.02.005}{\emph{Nucl. Phys. B}
  {\bfseries 893} (2015) 122}
  [\href{https://arxiv.org/abs/1411.5478}{{\ttfamily arXiv:1411.5478}}].

\bibitem{DiBari:2015oca}
P.~Di~Bari and S.~F. King, \emph{{Successful $N_2$ leptogenesis with flavour
  coupling effects in realistic unified models}},
  \href{https://doi.org/10.1088/1475-7516/2015/10/008}{\emph{JCAP} {\bfseries
  10} (2015) 008} [\href{https://arxiv.org/abs/1507.06431}{{\ttfamily
  arXiv:1507.06431}}].

\bibitem{DiBari:2017uka}
P.~Di~Bari and M.~Re~Fiorentin, \emph{{A full analytic solution of
  $SO(10)$-inspired leptogenesis}},
  \href{https://doi.org/10.1007/JHEP10(2017)029}{\emph{JHEP} {\bfseries 10}
  (2017) 029} [\href{https://arxiv.org/abs/1705.01935}{{\ttfamily
  arXiv:1705.01935}}].

\bibitem{Chianese:2018rnq}
M.~Chianese and P.~Di~Bari, \emph{{Strong thermal $SO(10)$-inspired
  leptogenesis in the light of recent results from long-baseline neutrino
  experiments}}, \href{https://doi.org/10.1007/JHEP05(2018)073}{\emph{JHEP}
  {\bfseries 05} (2018) 073}
  [\href{https://arxiv.org/abs/1802.07690}{{\ttfamily arXiv:1802.07690}}].

\bibitem{DiBari:2020plh}
P.~Di~Bari and R.~Samanta, \emph{{The $SO(10)$-inspired leptogenesis timely
  opportunity}}, \href{https://doi.org/10.1007/JHEP08(2020)124}{\emph{JHEP}
  {\bfseries 08} (2020) 124}
  [\href{https://arxiv.org/abs/2005.03057}{{\ttfamily arXiv:2005.03057}}].

\bibitem{DiBari:2005st}
P.~Di~Bari, \emph{{Seesaw geometry and leptogenesis}},
  \href{https://doi.org/10.1016/j.nuclphysb.2005.08.032}{\emph{Nucl. Phys. B}
  {\bfseries 727} (2005) 318}
  [\href{https://arxiv.org/abs/hep-ph/0502082}{{\ttfamily
  arXiv:hep-ph/0502082}}].

\bibitem{Vives:2005ra}
O.~Vives, \emph{{Flavor dependence of CP asymmetries and thermal leptogenesis
  with strong right-handed neutrino mass hierarchy}},
  \href{https://doi.org/10.1103/PhysRevD.73.073006}{\emph{Phys. Rev. D}
  {\bfseries 73} (2006) 073006}
  [\href{https://arxiv.org/abs/hep-ph/0512160}{{\ttfamily
  arXiv:hep-ph/0512160}}].

\bibitem{Fong:2014gea}
C.~S. Fong, D.~Meloni, A.~Meroni and E.~Nardi, \emph{{Leptogenesis in SO(10)}},
  \href{https://doi.org/10.1007/JHEP01(2015)111}{\emph{JHEP} {\bfseries 01}
  (2015) 111} [\href{https://arxiv.org/abs/1412.4776}{{\ttfamily
  arXiv:1412.4776}}].

\bibitem{Patel:2022xxu}
K.~M. Patel, \emph{{Minimal spontaneous CP-violating GUT and predictions for
  leptonic CP phases}},
  \href{https://doi.org/10.1103/PhysRevD.107.075041}{\emph{Phys. Rev. D}
  {\bfseries 107} (2023) 075041}
  [\href{https://arxiv.org/abs/2212.04095}{{\ttfamily arXiv:2212.04095}}].

\bibitem{NUFIT}
\emph{Nufit webpage, available online: \url{http://www.nu-fit.org/?q=node/238}
  (october 2021 data)}, .

\bibitem{Fong:2021xmi}
C.~S. Fong, \emph{{Cosmic evolution of lepton flavor charges}},
  \href{https://doi.org/10.1103/PhysRevD.105.043004}{\emph{Phys. Rev. D}
  {\bfseries 105} (2022) 043004}
  [\href{https://arxiv.org/abs/2109.04478}{{\ttfamily arXiv:2109.04478}}].

\bibitem{Buchmuller:2001sr}
W.~Buchmuller and M.~Plumacher, \emph{{Spectator processes and baryogenesis}},
  \href{https://doi.org/10.1016/S0370-2693(01)00614-1}{\emph{Phys. Lett. B}
  {\bfseries 511} (2001) 74}
  [\href{https://arxiv.org/abs/hep-ph/0104189}{{\ttfamily
  arXiv:hep-ph/0104189}}].

\bibitem{Nardi:2005hs}
E.~Nardi, Y.~Nir, J.~Racker and E.~Roulet, \emph{{On Higgs and sphaleron
  effects during the leptogenesis era}},
  \href{https://doi.org/10.1088/1126-6708/2006/01/068}{\emph{JHEP} {\bfseries
  01} (2006) 068} [\href{https://arxiv.org/abs/hep-ph/0512052}{{\ttfamily
  arXiv:hep-ph/0512052}}].

\bibitem{GellMann:1980vs}
M.~Gell-Mann, P.~Ramond and R.~Slansky, \emph{{Complex Spinors and Unified
  Theories}}, {\emph{Conf. Proc. C} {\bfseries 790927} (1979) 315}
  [\href{https://arxiv.org/abs/1306.4669}{{\ttfamily arXiv:1306.4669}}].

\bibitem{Garbrecht:2013bia}
B.~Garbrecht, F.~Glowna and P.~Schwaller, \emph{{Scattering Rates For
  Leptogenesis: Damping of Lepton Flavour Coherence and Production of Singlet
  Neutrinos}},
  \href{https://doi.org/10.1016/j.nuclphysb.2013.08.020}{\emph{Nucl. Phys. B}
  {\bfseries 877} (2013) 1} [\href{https://arxiv.org/abs/1303.5498}{{\ttfamily
  arXiv:1303.5498}}].

\bibitem{Garbrecht:2014kda}
B.~Garbrecht and P.~Schwaller, \emph{{Spectator Effects during Leptogenesis in
  the Strong Washout Regime}},
  \href{https://doi.org/10.1088/1475-7516/2014/10/012}{\emph{JCAP} {\bfseries
  10} (2014) 012} [\href{https://arxiv.org/abs/1404.2915}{{\ttfamily
  arXiv:1404.2915}}].

\bibitem{DOnofrio:2014rug}
M.~D'Onofrio, K.~Rummukainen and A.~Tranberg, \emph{{Sphaleron Rate in the
  Minimal Standard Model}},
  \href{https://doi.org/10.1103/PhysRevLett.113.141602}{\emph{Phys. Rev. Lett.}
  {\bfseries 113} (2014) 141602}
  [\href{https://arxiv.org/abs/1404.3565}{{\ttfamily arXiv:1404.3565}}].

\bibitem{Fong:2015vna}
C.~S. Fong, \emph{{Baryogenesis from Symmetry Principle}},
  \href{https://doi.org/10.1016/j.physletb.2015.11.055}{\emph{Phys. Lett. B}
  {\bfseries 752} (2016) 247}
  [\href{https://arxiv.org/abs/1508.03648}{{\ttfamily arXiv:1508.03648}}].

\bibitem{Blanchet:2011xq}
S.~Blanchet, P.~Di~Bari, D.~A. Jones and L.~Marzola, \emph{{Leptogenesis with
  heavy neutrino flavours: from density matrix to Boltzmann equations}},
  \href{https://doi.org/10.1088/1475-7516/2013/01/041}{\emph{JCAP} {\bfseries
  01} (2013) 041} [\href{https://arxiv.org/abs/1112.4528}{{\ttfamily
  arXiv:1112.4528}}].

\bibitem{Fong:2020fwk}
C.~S. Fong, \emph{{Baryogenesis in the Standard Model and its Supersymmetric
  Extension}}, \href{https://doi.org/10.1103/PhysRevD.103.L051705}{\emph{Phys.
  Rev. D} {\bfseries 103} (2021) L051705}
  [\href{https://arxiv.org/abs/2012.03973}{{\ttfamily arXiv:2012.03973}}].

\bibitem{Salvio:2011sf}
A.~Salvio, P.~Lodone and A.~Strumia, \emph{{Towards leptogenesis at NLO: the
  right-handed neutrino interaction rate}},
  \href{https://doi.org/10.1007/JHEP08(2011)116}{\emph{JHEP} {\bfseries 08}
  (2011) 116} [\href{https://arxiv.org/abs/1106.2814}{{\ttfamily
  arXiv:1106.2814}}].

\bibitem{Laine:2011pq}
M.~Laine and Y.~Schroder, \emph{{Thermal right-handed neutrino production rate
  in the non-relativistic regime}},
  \href{https://doi.org/10.1007/JHEP02(2012)068}{\emph{JHEP} {\bfseries 02}
  (2012) 068} [\href{https://arxiv.org/abs/1112.1205}{{\ttfamily
  arXiv:1112.1205}}].

\bibitem{Garbrecht:2019zaa}
B.~Garbrecht, P.~Klose and C.~Tamarit, \emph{{Relativistic and spectator
  effects in leptogenesis with heavy sterile neutrinos}},
  \href{https://doi.org/10.1007/JHEP02(2020)117}{\emph{JHEP} {\bfseries 02}
  (2020) 117} [\href{https://arxiv.org/abs/1904.09956}{{\ttfamily
  arXiv:1904.09956}}].

\bibitem{Antusch:2005gp}
S.~Antusch, J.~Kersten, M.~Lindner, M.~Ratz and M.~A. Schmidt, \emph{{Running
  neutrino mass parameters in see-saw scenarios}},
  \href{https://doi.org/10.1088/1126-6708/2005/03/024}{\emph{JHEP} {\bfseries
  03} (2005) 024} [\href{https://arxiv.org/abs/hep-ph/0501272}{{\ttfamily
  arXiv:hep-ph/0501272}}].

\bibitem{Super-Kamiokande:2020wjk}
{\scshape Super-Kamiokande} collaboration, \emph{{Search for proton decay via
  $p\to e^+\pi^0$ and $p\to \mu^+\pi^0$ with an enlarged fiducial volume in
  Super-Kamiokande I-IV}},
  \href{https://doi.org/10.1103/PhysRevD.102.112011}{\emph{Phys. Rev. D}
  {\bfseries 102} (2020) 112011}
  [\href{https://arxiv.org/abs/2010.16098}{{\ttfamily arXiv:2010.16098}}].

\bibitem{Planck:2018vyg}
{\scshape Planck} collaboration, \emph{{Planck 2018 results. VI. Cosmological
  parameters}},
  \href{https://doi.org/10.1051/0004-6361/201833910}{\emph{Astron. Astrophys.}
  {\bfseries 641} (2020) A6}
  [\href{https://arxiv.org/abs/1807.06209}{{\ttfamily arXiv:1807.06209}}].

\bibitem{Antusch:2013jca}
S.~Antusch and V.~Maurer, \emph{{Running quark and lepton parameters at various
  scales}}, \href{https://doi.org/10.1007/JHEP11(2013)115}{\emph{JHEP}
  {\bfseries 11} (2013) 115} [\href{https://arxiv.org/abs/1306.6879}{{\ttfamily
  arXiv:1306.6879}}].

\bibitem{Esteban:2020cvm}
I.~Esteban, M.~C. Gonzalez-Garcia, M.~Maltoni, T.~Schwetz and A.~Zhou,
  \emph{{The fate of hints: updated global analysis of three-flavor neutrino
  oscillations}}, \href{https://doi.org/10.1007/JHEP09(2020)178}{\emph{JHEP}
  {\bfseries 09} (2020) 178}
  [\href{https://arxiv.org/abs/2007.14792}{{\ttfamily arXiv:2007.14792}}].

\bibitem{Davidson:2002qv}
S.~Davidson and A.~Ibarra, \emph{{A Lower bound on the right-handed neutrino
  mass from leptogenesis}},
  \href{https://doi.org/10.1016/S0370-2693(02)01735-5}{\emph{Phys. Lett. B}
  {\bfseries 535} (2002) 25}
  [\href{https://arxiv.org/abs/hep-ph/0202239}{{\ttfamily
  arXiv:hep-ph/0202239}}].

\bibitem{Antusch:2023zjk}
S.~Antusch, K.~Hinze, S.~Saad and J.~Steiner, \emph{{Singling out SO(10) GUT
  models using recent PTA results}},
  \href{https://doi.org/10.1103/PhysRevD.108.095053}{\emph{Phys. Rev. D}
  {\bfseries 108} (2023) 095053}
  [\href{https://arxiv.org/abs/2307.04595}{{\ttfamily arXiv:2307.04595}}].

\bibitem{Xu:2023wog}
H.~Xu et~al., \emph{{Searching for the Nano-Hertz Stochastic Gravitational Wave
  Background with the Chinese Pulsar Timing Array Data Release I}},
  \href{https://doi.org/10.1088/1674-4527/acdfa5}{\emph{Res. Astron.
  Astrophys.} {\bfseries 23} (2023) 075024}
  [\href{https://arxiv.org/abs/2306.16216}{{\ttfamily arXiv:2306.16216}}].

\bibitem{EPTA:2023fyk}
{\scshape EPTA, InPTA:} collaboration, \emph{{The second data release from the
  European Pulsar Timing Array - III. Search for gravitational wave signals}},
  \href{https://doi.org/10.1051/0004-6361/202346844}{\emph{Astron. Astrophys.}
  {\bfseries 678} (2023) A50}
  [\href{https://arxiv.org/abs/2306.16214}{{\ttfamily arXiv:2306.16214}}].

\bibitem{NANOGrav:2023gor}
{\scshape NANOGrav} collaboration, \emph{{The NANOGrav 15 yr Data Set: Evidence
  for a Gravitational-wave Background}},
  \href{https://doi.org/10.3847/2041-8213/acdac6}{\emph{Astrophys. J. Lett.}
  {\bfseries 951} (2023) L8}
  [\href{https://arxiv.org/abs/2306.16213}{{\ttfamily arXiv:2306.16213}}].

\bibitem{Reardon:2023gzh}
D.~J. Reardon et~al., \emph{{Search for an Isotropic Gravitational-wave
  Background with the Parkes Pulsar Timing Array}},
  \href{https://doi.org/10.3847/2041-8213/acdd02}{\emph{Astrophys. J. Lett.}
  {\bfseries 951} (2023) L6}
  [\href{https://arxiv.org/abs/2306.16215}{{\ttfamily arXiv:2306.16215}}].

\bibitem{NANOGrav:2023hvm}
{\scshape NANOGrav} collaboration, \emph{{The NANOGrav 15 yr Data Set: Search
  for Signals from New Physics}},
  \href{https://doi.org/10.3847/2041-8213/acdc91}{\emph{Astrophys. J. Lett.}
  {\bfseries 951} (2023) L11}
  [\href{https://arxiv.org/abs/2306.16219}{{\ttfamily arXiv:2306.16219}}].

\bibitem{Santos:2024eko}
{\scshape Super-Kamiokande} collaboration, \emph{{Latest results from
  Super-Kamiokande}},  in \emph{{58th Rencontres de Moriond on Electroweak
  Interactions and Unified Theories}}, 5, 2024,
  \href{https://arxiv.org/abs/2405.07900}{{\ttfamily arXiv:2405.07900}}.

\bibitem{DESI:2024mwx}
{\scshape DESI} collaboration, \emph{{DESI 2024 VI: Cosmological Constraints
  from the Measurements of Baryon Acoustic Oscillations}},
  \href{https://arxiv.org/abs/2404.03002}{{\ttfamily arXiv:2404.03002}}.

\bibitem{Dev:2022jbf}
P.~S.~B. Dev et~al., \emph{{Searches for Baryon Number Violation in Neutrino
  Experiments: A White Paper}},
  \href{https://arxiv.org/abs/2203.08771}{{\ttfamily arXiv:2203.08771}}.

\bibitem{FileviezPerez:2004hn}
P.~Fileviez~Perez, \emph{{Fermion mixings versus d = 6 proton decay}},
  \href{https://doi.org/10.1016/j.physletb.2004.06.061}{\emph{Phys. Lett.}
  {\bfseries B595} (2004) 476}
  [\href{https://arxiv.org/abs/hep-ph/0403286}{{\ttfamily
  arXiv:hep-ph/0403286}}].

\bibitem{Nath:2006ut}
P.~Nath and P.~Fileviez~Perez, \emph{{Proton stability in grand unified
  theories, in strings and in branes}},
  \href{https://doi.org/10.1016/j.physrep.2007.02.010}{\emph{Phys. Rept.}
  {\bfseries 441} (2007) 191}
  [\href{https://arxiv.org/abs/hep-ph/0601023}{{\ttfamily
  arXiv:hep-ph/0601023}}].

\bibitem{Aoki:2017puj}
Y.~Aoki, T.~Izubuchi, E.~Shintani and A.~Soni, \emph{{Improved lattice
  computation of proton decay matrix elements}},
  \href{https://doi.org/10.1103/PhysRevD.96.014506}{\emph{Phys. Rev. D}
  {\bfseries 96} (2017) 014506}
  [\href{https://arxiv.org/abs/1705.01338}{{\ttfamily arXiv:1705.01338}}].

\bibitem{Nihei:1994tx}
T.~Nihei and J.~Arafune, \emph{{The Two loop long range effect on the proton
  decay effective Lagrangian}},
  \href{https://doi.org/10.1143/PTP.93.665}{\emph{Prog. Theor. Phys.}
  {\bfseries 93} (1995) 665}
  [\href{https://arxiv.org/abs/hep-ph/9412325}{{\ttfamily
  arXiv:hep-ph/9412325}}].

\bibitem{Buras:1977yy}
A.~J. Buras, J.~R. Ellis, M.~K. Gaillard and D.~V. Nanopoulos, \emph{{Aspects
  of the Grand Unification of Strong, Weak and Electromagnetic Interactions}},
  \href{https://doi.org/10.1016/0550-3213(78)90214-6}{\emph{Nucl. Phys. B}
  {\bfseries 135} (1978) 66}.

\bibitem{Goldman:1980ah}
J.~T. Goldman and D.~A. Ross, \emph{{How Accurately Can We Estimate the Proton
  Lifetime in an SU(5) Grand Unified Model?}},
  \href{https://doi.org/10.1016/0550-3213(80)90371-5}{\emph{Nucl. Phys. B}
  {\bfseries 171} (1980) 273}.

\bibitem{Caswell:1982fx}
W.~E. Caswell, J.~Milutinovic and G.~Senjanovic, \emph{{Predictions of
  Left-right Symmetric Grand Unified Theories}},
  \href{https://doi.org/10.1103/PhysRevD.26.161}{\emph{Phys. Rev. D} {\bfseries
  26} (1982) 161}.

\bibitem{Ibanez:1984ni}
L.~E. Ibanez and C.~Munoz, \emph{{Enhancement Factors for Supersymmetric Proton
  Decay in the {Wess-Zumino} Gauge}},
  \href{https://doi.org/10.1016/0550-3213(84)90439-5}{\emph{Nucl. Phys. B}
  {\bfseries 245} (1984) 425}.

\bibitem{Abbott:1980zj}
L.~F. Abbott and M.~B. Wise, \emph{{The Effective Hamiltonian for Nucleon
  Decay}}, \href{https://doi.org/10.1103/PhysRevD.22.2208}{\emph{Phys. Rev. D}
  {\bfseries 22} (1980) 2208}.

\bibitem{Babu:2015bna}
K.~S. Babu and S.~Khan, \emph{{Minimal nonsupersymmetric $SO(10)$ model: Gauge
  coupling unification, proton decay, and fermion masses}},
  \href{https://doi.org/10.1103/PhysRevD.92.075018}{\emph{Phys. Rev. D}
  {\bfseries 92} (2015) 075018}
  [\href{https://arxiv.org/abs/1507.06712}{{\ttfamily arXiv:1507.06712}}].

\bibitem{Chakrabortty:2019fov}
J.~Chakrabortty, R.~Maji and S.~F. King, \emph{{Unification, Proton Decay and
  Topological Defects in non-SUSY GUTs with Thresholds}},
  \href{https://doi.org/10.1103/PhysRevD.99.095008}{\emph{Phys. Rev. D}
  {\bfseries 99} (2019) 095008}
  [\href{https://arxiv.org/abs/1901.05867}{{\ttfamily arXiv:1901.05867}}].

\bibitem{Babu:2024ecl}
K.~S. Babu, B.~Bajc and V.~Susi\v{c}, \emph{{A realistic theory of E$_{6}$
  unification through novel intermediate symmetries}},
  \href{https://doi.org/10.1007/JHEP06(2024)018}{\emph{JHEP} {\bfseries 06}
  (2024) 018} [\href{https://arxiv.org/abs/2403.20278}{{\ttfamily
  arXiv:2403.20278}}].

\bibitem{Georgi:1979md}
H.~Georgi, \emph{{Towards a Grand Unified Theory of Flavor}},
  \href{https://doi.org/10.1016/0550-3213(79)90497-8}{\emph{Nucl. Phys. B}
  {\bfseries 156} (1979) 126}.

\bibitem{delAguila:1980qag}
F.~del Aguila and L.~E. Ibanez, \emph{{Higgs Bosons in SO(10) and Partial
  Unification}},
  \href{https://doi.org/10.1016/0550-3213(81)90266-2}{\emph{Nucl. Phys. B}
  {\bfseries 177} (1981) 60}.

\bibitem{Mohapatra:1982aq}
R.~N. Mohapatra and G.~Senjanovic, \emph{{Higgs Boson Effects in Grand Unified
  Theories}}, \href{https://doi.org/10.1103/PhysRevD.27.1601}{\emph{Phys. Rev.
  D} {\bfseries 27} (1983) 1601}.

\bibitem{Dimopoulos:1984ha}
S.~Dimopoulos and H.~M. Georgi, \emph{{Extended Survival Hypothesis and Fermion
  Masses}}, \href{https://doi.org/10.1016/0370-2693(84)91049-9}{\emph{Phys.
  Lett. B} {\bfseries 140} (1984) 67}.

\bibitem{Saad:2017pqj}
S.~Saad, \emph{{Fermion Masses and Mixings, Leptogenesis and Baryon Number
  Violation in Pati-Salam Model}},
  \href{https://doi.org/10.1016/j.nuclphysb.2019.114630}{\emph{Nucl. Phys. B}
  {\bfseries 943} (2019) 114630}
  [\href{https://arxiv.org/abs/1712.04880}{{\ttfamily arXiv:1712.04880}}].

\bibitem{Machacek:1983tz}
M.~E. Machacek and M.~T. Vaughn, \emph{{Two Loop Renormalization Group
  Equations in a General Quantum Field Theory. 1. Wave Function
  Renormalization}},
  \href{https://doi.org/10.1016/0550-3213(83)90610-7}{\emph{Nucl. Phys. B}
  {\bfseries 222} (1983) 83}.

\bibitem{Weinberg:1980wa}
S.~Weinberg, \emph{{Effective Gauge Theories}},
  \href{https://doi.org/10.1016/0370-2693(80)90660-7}{\emph{Phys. Lett. B}
  {\bfseries 91} (1980) 51}.

\bibitem{Hall:1980kf}
L.~J. Hall, \emph{{Grand Unification of Effective Gauge Theories}},
  \href{https://doi.org/10.1016/0550-3213(81)90498-3}{\emph{Nucl. Phys. B}
  {\bfseries 178} (1981) 75}.

\bibitem{Bertolini:2009qj}
S.~Bertolini, L.~Di~Luzio and M.~Malinsky, \emph{{Intermediate mass scales in
  the non-supersymmetric SO(10) grand unification: A Reappraisal}},
  \href{https://doi.org/10.1103/PhysRevD.80.015013}{\emph{Phys. Rev. D}
  {\bfseries 80} (2009) 015013}
  [\href{https://arxiv.org/abs/0903.4049}{{\ttfamily arXiv:0903.4049}}].

\bibitem{Bertolini:2013vta}
S.~Bertolini, L.~Di~Luzio and M.~Malinsky, \emph{{Light color octet scalars in
  the minimal SO(10) grand unification}},
  \href{https://doi.org/10.1103/PhysRevD.87.085020}{\emph{Phys. Rev. D}
  {\bfseries 87} (2013) 085020}
  [\href{https://arxiv.org/abs/1302.3401}{{\ttfamily arXiv:1302.3401}}].

\bibitem{Hyper-Kamiokande:2018ofw}
{\scshape Hyper-Kamiokande} collaboration, \emph{{Hyper-Kamiokande Design
  Report}},  \href{https://arxiv.org/abs/1805.04163}{{\ttfamily
  arXiv:1805.04163}}.

\bibitem{KamLAND-Zen:2022tow}
{\scshape KamLAND-Zen} collaboration, \emph{{Search for the Majorana Nature of
  Neutrinos in the Inverted Mass Ordering Region with KamLAND-Zen}},
  \href{https://doi.org/10.1103/PhysRevLett.130.051801}{\emph{Phys. Rev. Lett.}
  {\bfseries 130} (2023) 051801}
  [\href{https://arxiv.org/abs/2203.02139}{{\ttfamily arXiv:2203.02139}}].

\bibitem{nEXO:2021ujk}
{\scshape nEXO} collaboration, \emph{{nEXO: neutrinoless double beta decay
  search beyond 10$^{28}$ year half-life sensitivity}},
  \href{https://doi.org/10.1088/1361-6471/ac3631}{\emph{J. Phys. G} {\bfseries
  49} (2022) 015104} [\href{https://arxiv.org/abs/2106.16243}{{\ttfamily
  arXiv:2106.16243}}].

\end{thebibliography}\endgroup
\end{document}